\documentclass[useAMS,usenatbib]{mn2e}
\usepackage{graphicx,amsmath}
\usepackage{amssymb}
\usepackage{wrapfig}
\usepackage[british,english]{babel}
\usepackage{times}

\title[The size of the Comptonizing medium]{Constraining the size of the 
Comptonizing medium by modeling
the energy dependent time-lags of kHz QPOs of Neutron star system}

\author[Kumar and Misra]{Nagendra Kumar$^{1}$\thanks{E-mail:nagendrak@iucaa.in} and Ranjeev Misra$^{1}$
\thanks{E-mail: rmisra@iucaa.in}\\
$^{1}$\textit{Inter-University Centre For Astronomy and Astrophysics, Post Bag4, Ganeshkind, 
Pune-411007, India }}

\begin{document}
\date{}

\pagerange{\pageref{firstpage}--\pageref{lastpage}} 
\maketitle

\label{firstpage}

\begin{abstract}
In earlier works, we had shown that the observed soft lags and
r.m.s versus energy of the lower kHz QPO of neutron star binaries 
can be explained in the framework of a thermal Comptonization model.
It was also shown that such an interpretation can provide estimates
of the size and geometry of the Comptonizing medium. Here we study
the dependence of these estimates on the time-averaged spectral
model assumed and on the frequency of the QPO. We use the high quality
time lag and r.m.s obtained during March 3rd 1996 observation of
4U 1608-52 by RXTE as well as other observations of the source
at different QPO frequencies where a single time-lag between 
two broad energy bands have been reported. We compare the results
obtained when assuming that the time-averaged spectra are represented
by the spectrally degenerate ``hot'' and ``cold'' seed photon spectral models.
We find that for the ``hot'' seed photon model the medium size is
in the range of 0.3-2.0 kms and the size decreases with increasing 
QPO frequency. On the other hand for the ``cold'' seed photon model
the range for the sizes are much larger 0.5-20 kms and hence perhaps show
no variation with QPO frequency. Our results emphasis the need for broad
band spectral information combined with high frequency timing to lift
this degeneracy. We further show that the r.m.s as a function of energy
for the upper kHz QPO is similar to the lower one and indeed we
find that the driver for this QPO should be temperature variations of
the corona identical to the lower kHz QPO. However, the time lag
reported for the upper kHz QPO is hard, which if confirmed, would
challenge the simple Comptonization model presented here. It would
perhaps imply that reverberation lags are also important and/or the dominate
spectral component is not a single temperature medium but a multi-temperature
complex one.

\end{abstract}


\section{Introduction}
In neutron star  low-mass X-ray binaries (LMXBs) systems, the neutron star
accretes  matter via Roche lobe overflow from a normal low-mass companion 
star. This accretion forms  a disk
and it is believed that the intense X-ray emissions arise from such an 
accretion disk.
The X-rays can be emitted from the inner accretion disk and/or the boundary
layer between the neutron star  surface and the disk.     
These LMXBs are, in general, weakly magnetised, non-pulsating systems
and are broadly divided phenomenologically into two sub-classes depending 
upon their long term X-ray luminosity 
variations. Some sources are considered to be `transient' if their
 luminosities changes by more  than 100 times while other are considered
`persistent' if their  luminosities varies by  factors of $\sim$ 2--10 and if
they have always been detectable over the history of X-ray astronomy 
\citep{Psaltis2006}. 
Further, they are divided into two classes `Z', and `atoll',
based on the correlation between their X-ray spectral and timing properties  \citep{Hasinger-vanderKlis1989}. In the color-color diagrams (CCDs), 
the Z sources trace out a roughly Z-shaped track within hours/day. 
The track of  atoll sources in the  CCD are generally C-shaped and the source
 covers the track 
 on time scales of days to years. 
In general, all Z sources are persistent while atoll sources can be either
 persistent or  transient. However, the classification may depend only on the
luminosity since XTE J1701-462 shows characteristics  of 
both atoll and Z sources depending on its flux level 
\citep{Lin-Remillard-Homan2009,Homan-etal2010}.

Quasi-periodic oscillations (QPO) in the frequency range, 400 -- 1300
Hz have been discovered in most neutron star LMXBs with the Rossi
X-ray Timing Explorer (RXTE) satellite \citep{vanderKlis2006a}. Often
two simultaneous kHz QPO  are observed with their frequency separation
ranging from $\sim 200$ to $\sim 400$ Hz and with their frequencies 
correlated with each other
\citep{Belloni-Mendez-Homan2005,Belloni-Mendez-Homan2007}.  The QPO
with the higher frequency is referred to as the ``upper'' kHz QPO while
the other is called the ``lower'' kHz QPO.

A number of models have been proposed to physically interpret these
twin kHz QPOs. Some of these models are: the sonic-point model
\citep{Miller-Lamb-Psaltis1998,Lamb-Miller2003}; the relativistic
precession model  \citep{Stella-Vietri1999}; the two-oscillator model
\citep{ Osherovich-Titarchuk1999,Titarchuk2003}; the relativistic
resonant model  \citep{Kluzniak-Abramowicz2001}; deformed-disk
oscillation model  \citep{Kato2009,Mukhopadhyay2009}); and magneto-hydrodynamic
 (MHD) models \citep{Zhang2004,Li-Zhang2005,
Shi-Li2009,Erkut-Psaltis-Alpar2008}. A particular model is favoured if
it can describe the correlation between the kHz frequencies with the properties 
of the lower frequency QPOs (10 -- 200 Hz) which are also
observed in these sources
\citep[e.g.][]{Straaten-vanderKlis-Mendez2003,Altamirano-etal2008}. A
model may  also need to explain the conditions which determine  the
appearance of  twin versus a single kHz QPO
\cite[e.g.][]{Sanna-etal2012,Lin-etal2012,Mendez-etal1998b}.
Unfortunately, there is no consensus on which among these is the  most
likely model \cite[e.g.][]{Lin-etal2011,Wang-etal2013}.

In this scenario, it maybe more prudent to understand the relation between
the kHz QPOs and the X-ray flux and spectra and to identify the radiative
process responsible for the QPOs irrespective of their dynamical origin. 
The frequencies of the kHz QPOs exhibit complex correlation with
intensity. In short time-scales of hours they are positively correlated
\citep{Yu-etal1997,Ford-etal2000}, while
on longer time-scales they exhibit  parallel tracks in the kHz QPO frequency 
versus intensity diagram \citep{Mendez-etal1999}. The incidence of kHz QPOs
also depends on the X-ray intensity \citep{Misra-Shanthi2004}.
The frequencies depend on the position of the source in the
color-color diagram and two different tracks are traced out by the
higher and lower kHz QPOS. 
Whenever a single kHz QPO is observed, it follow either the upper or
lower kHz QPOs track \citep{Mendez-vanderKlis1999,vanderKlis2000}. In
general, the occurrence of kHz QPOs on  different positions of CCDs
indicates  that as the QPO frequency decreases, the X-ray spectrum
tends to be  harder, i.e., its hard color increases
\citep{Straaten-vanderKlis-Mendez2003,Altamirano-etal2008,Lin-etal2012}.

The fractional root mean square (r.m.s.) amplitude of the kHz QPOs
usually increases with photons energy at least up to 20 keV
\citep{Berger-etal1996,Zhang-etal1996,Wijnands-etal1997b,vanderKlis2006a}.
Beyond 20 keV the dependence is unknown with some indications that
it may decrease 
\citep{Mukherjee-Bhattacharyya2012}.  In general, the strength of
lower kHz QPOs is larger than the upper one at any energy bin
\citep{Mendez-etal1998a,Wijnands-etal1997a}.
Another important characteristic is the energy dependent time-lag
shown by these QPOs.   
It was found  that the $\sim$18 keV photons have a time lag of
$\sim -50 \mu s$
with respect to the $\sim$5 keV photons in the
 $\sim$850 Hz lower kHz QPO in 4U 1608-52, the soft photon came later
and hence these lags are referred to as ``soft lag''.
The soft lag associated with the lower kHz QPO was also
found for another  source 4U 1636-53 \citep{Vaughan-etal1997,
Kaaret-etal1999}. Recently \cite{Peille-Barret-Uttley2015}, \cite{Barret2013},
and \cite{deAvellar-etal2013} have   confirmed the soft lag in lower kHz
QPOs by analysing a large number of  observations and found that
the magnitude of the time-lags have a complex dependence on the
QPO frequency.
 Moreover, \cite{deAvellar-etal2013}
claimed that the time lag of the lower and the upper kHz QPOs have
opposite signs, i.e., the upper  kHz QPOs exhibit  hard lags.

The time averaged X-ray spectral modelling of NS LMXBs reveal  that in these 
system, the inner region is partially
or fully covered  by a corona consisting of hot thermal electrons which
Compton upscatter low energy photons.
There are two types of phenomenological spectral  models which are used to
 describe the X-ray emission in NS LMXBs. 
Both models describe the X-ray spectrum as sum of a soft black body like
component and the other a harder component due to thermal Comptonization.
In one model, 
the thermal component is interpreted as a  multicolor disk black body (MCD) 
emission attributed to accretion disk, while the seed photons for the 
Comptonized component are emitted from close to the NS surface or the 
boundary layer \citep{Mitsuda-etal1984,Mitsuda-etal1989}.
In the other  model \citep{White-etal1986}, the soft component is a 
single temperature black body emission from the boundary layer,
 while the seed photons for the 
Comptonization are emitted from the accretion disk 
\cite[see for e.g.][for reviews of LMXBs X-ray spectral models]{
Barret2001,Salvo-stella2002,Paizis-etal2006}. 
These two approaches of 
modelling are often degenerate i.e. they give equally 
statistically acceptable fit to data. Apart from the theoretical
differences the critical radiative spectral difference between the
two models is that in one  the temperature of the direct soft component 
spectrum  is  lower \cite[e.g.][]{Gierlinski-Done2002, 
Tarana-Bazzano-Ubertini2008,Agrawal-misra2009,Raichur-Misra-Dewangan2011,
Barret2013} and in the other higher \cite[e.g.][]{Agrawal-Sreekumar2003,Paizis-etal2005, 
Farinelli-Titarchuk-Frontera2007,Farinelli-etal2008} than that of the 
seed photon source temperature of the thermal Comptonization. Thus in this work,
following 
\cite{Lin-Remillard-Homan2007}, we refer to the former spectral model 
as ``hot-seed model'' and the latter model as 
``cold-seed model''. The typical spectral parameter values varies
depending on the spectral state and in general, 
from the soft  to the hard state,  the temperature of corona ranges 
from $\sim 2$ keV to $\sim 20$ keV and the optical depth, $\tau_o$ 
ranges from $\sim$ 10 to $\sim$ 2. 

Energy dependent r.m.s and time lags of the kHz QPOs provide information
on the spectral parameter whose variation is responsible for the
phenomenon and can constrain the size and geometry of the system
\citep{Lee-Miller1998}. This is also true for other lower frequency temporal
behaviour as for example, in black hole X-ray binary system, 
the positive correlation between the 
fractional r.m.s. and photons energy has been explained due to corona 
temperature oscillation \citep{Gierlinski-Zdziarski2005}. For
thermal Comptonization, since the higher energy photons scatter more than the 
soft ones, it is expected that the system should show  hard lag which
is contradictory to the observations of kHz QPOs.
 However, \cite{Lee-Misra-Taam2001} have shown that a system 
will show soft lags if some fraction of the
Comptonized photons impinge back to the input source. There
could also be time-lags introduced due to reflection of the
X-rays from an accretion disk \citep{Barret2013}.

In \cite{Kumar-Misra2014} (hereafter Paper I) we studied the expected 
time lag due to thermal Comptonization by solving the linearised time
dependent Kompaneets equation for different physical situations such
as the primary oscillation being in the soft photon source or in the
heating rate of the corona. We then compared the results with the
RXTE observations of 4U 1608-52 on 3 March 1996 and concluded that
the model can explain the data and inferred the corona's size to be
0.3--2 km. The results depend on the steady state spectral
parameters and  there we had used a particular fitting of the data.
However, as mentioned above, the spectral parameters of LMXBs 
in general are typically degenerate and it is important to check 
the differences obtained when one uses the cold-seed photon model 
or the hot-seed photon one. Moreover, the  results of
\cite{Barret2013} and \cite{deAvellar-etal2013} allow us to compare
the model predictions for different kHz QPO frequencies and hence
possibly infer the size and geometry as a function of frequency. 
Such a study will be useful to constrain hydrodynamical models which relate
the size of the corona to the kHz QPO frequency such as those given
by  \cite{Pal-Chakrabarti2014} and \cite{Cabanac-etal2010}.

In the next section we briefly review the model using the
 linearised time dependent Kompaneets equation which is extensively
described in Paper 1. In \S 3, we compare the model predictions for
the low frequency kHz QPO observed in 4U 1608-52 on 3rd March 1996 considering
 both the soft and hot seed photon spectral models. In \S 4 we discuss the
implication of the model on the high frequency kHz QPO. In \S5 we try to
constrain the size and geometry of the system for different frequencies,
while in \S6 we summarise and discuss the results.

\section{Variability of Thermal Comptonized photons}
In the non-relativistic limit (i.e. kT$_e$ $\ll$ m$_e c^2$) and for low
photon energies (E $\ll$ m$_e c^2$), the evolution of the photon density 
($n_\gamma$) inside a Comptonizing medium is governed by the Kompaneets 
equation \citep{Kompaneets1957},

\begin{align}
t_c\frac{dn_\gamma}{dt} =& \frac{1}{m_ec^2}\frac{d}{dE}
\left[-4kT_eEn_\gamma + E^2n_\gamma + kT_e \frac{d}{dE}(E^2n_\gamma)
\right]  \nonumber \\ &
+ t_c\dot{n}_{s\gamma} - t_c\dot{n}_{esc}
\end{align}
where the induced scattering term has been neglected
and the equation is written in terms of the photon density rather 
than photon occupation number. Here
\begin{equation}
\dot{n}_{s\gamma}= \left[\frac{4  \pi  a^2}{V_c}\right] 
\left( \frac{2\pi}{h^3c^2} \frac{E^2}{(\exp\left[{\frac{E}{kT_b}}\right]-1)}
\right)
\end{equation}
is the rate of input photons per unit volume.  A 
simplified geometry has been assumed in which a spherical
black body seed photon source at a temperature $T_b$ and radius $a$
 is surrounded by a corona of width $L$ and temperature $T_e$. Thus, 
the Comptonizing  medium has a volume  $V_c = (4/3) \pi [(a+L)^3-a^3]$, 
the Thompson collision time scale $t_c = 1/(c n_e \sigma_T)$, and the 
optical depth of  medium $\tau = \frac{(L/c)}{t_c}$, 
where $\sigma_T$ is the Thompson cross-section  and $n_e$ 
is the electron density.
The escape rate of the photon density is taken to be 
$\dot{n}_{esc} \simeq$ $\frac{n_\gamma}{(\tau^2+\tau)t_c}$, where 
$(\tau^2+\tau)$ is assumed to be the average number of scatterings. 
The emergent X-ray spectrum is determined by $\dot{n}_{esco}$ in 
steady state (i.e., 
$\frac{dn_\gamma}{dt} = 0$) where $n_{\gamma o}$ is computed for corresponding
steady state (or time-averaged) values of $T_{eo}$ and $T_{bo}$ and $\tau$ .

The corona temperature reaches a steady state when the external heating
rate per electron ($\dot{H}_{Ex}$, although its nature is unknown) is balanced 
by the Compton cooling rate per electron (${\langle \Delta \dot E \rangle} = 
\int_{E_{min}}^{E^{max}}(4kT_e-E) \frac{E}{m_ec^2}\ n_{\gamma} \ \sigma_T \ c\ 
dE$). In general the time evolution of corona temperature is described as :
\begin{equation}
\frac{3}{2}k\frac{\partial T_e}{\partial t} 
= \dot{H}_{Ex} - {\langle \Delta \dot E \rangle}
\label{Heat}
\end{equation}

The soft photon source has an internal heating rate which is taken
to be $4 \pi a^2 \sigma (T'_{b})^4$, such that in the absence of any another
heating it temperature would be $T'_b$. Since we consider the
possibility that a fraction $\eta$, of the Comptonized photons 
impinge back onto to the seed photon source, the actual temperature
$T_b$ is given by,
 \begin{equation} 
4 \pi a^2 \sigma T_{b}^4 =4 \pi a^2 \sigma T_{b}^{'4}  + \eta V_c \int 
\frac{n_{\gamma }}{(\tau^2+\tau)t_c} E \ dE 
\label{feedback}
\end{equation}
For a given value of $T_b$, there is a maximum allowable value of $\eta$, 
$\eta_{max}$ for which 
$T_{bo}^{'}$ is zero. The variation in $\dot{H}_{Ex}$ will lead to variation
in $T_e$ while $T_b$ can be varied from its steady state value due to fluctuations in the 
back-scattered photons. 
If their 
variability amplitude is small, the time averaged X-ray spectrum will 
correspond to the steady parameters, $T_{eo}$, $\tau$ and $T_{bo}$.

The energy dependent temporal features are assumed to be driven by fluctuation
of the medium temperature and/or the source photon temperature over their averaged 
values i.e. $T_e = T_{eo}(1+\Delta T_e \ e^{-i\omega t})$,
or $T_b = T_{bo} (1+ \Delta  T_b \ e^{-i\omega t})$. 
Here, $\omega$ is the angular frequency of the oscillation and their 
oscillation amplitude $\Delta T_e \ll 1$ and $\Delta T_b \ll 1$ are in general 
complex quantities. These fluctuation will lead to variation in photon density
$n_\gamma = n_{\gamma o}(1+ \Delta n_\gamma \ e^{-i\omega t})$ and its 
amplitude $\Delta n_\gamma$ is computed using the linearised Kompaneets equation\citep[e.g.][]{Lee-Miller1998,Lee-Misra-Taam2001}:
\begin{align}
-&\frac{d^2\Delta n_\gamma}{dE^2}+\left(\frac{-1}{kT_{eo}}-
\frac{2}{n_{\gamma o}}\frac{dn_{\gamma o}}{dE}\right)
\frac{d\Delta n_{\gamma}}{dE} \nonumber \\ &
+\frac{m_ec^2t_c(\dot{n}_{s\gamma o}-i\omega n_{\gamma o})}
{E^2n_{\gamma o}kT_{eo}}\Delta n_\gamma =  
\left(\frac{-2}{E^2}+\frac{1}{n_{\gamma o}}
\frac{d^2n_{\gamma o}}{dE^2}\right)\Delta T_e \nonumber \\ &+
\frac{m_ec^2t_c\dot{n}_{s\gamma o}}{E^2n_{\gamma o}kT_{eo}}
\left(\frac{\frac{E}{kT_{bo}}}{1-\exp{\left(\frac{-E}{kT_{bo}}
\right)}}\right)\Delta T_b
\label{Komlin}
\end{align}
$\Delta n_\gamma (E)$ determines the energy dependent temporal features. In
particular $|\Delta n_\gamma (E)|$ is the fractional r.m.s, and the argument of
$[\Delta n_\gamma(E_1)\Delta n^*_\gamma(E_2)]$ is the phase lag between two 
energies $E_1$ and $E_2$. The mean phase lag between two energy band (say, 
$(E_1 to E_2)$ $\&$ $(E_3 to E_4)$) is the argument of 
$[\Delta n_\gamma(E_{12})
\Delta n^*_\gamma(E_{34})]$,  where $\Delta n_\gamma(E_{ab}) = 
\frac{\int_{E_{a}}^{E^{b}} n_{\gamma o} \Delta n_\gamma dE}
{\int_{E_{a}}^{E^{b}} n_{\gamma o} dE}$.

Two possible cases are considered which can drive the temporal features of 
X-ray spectrum. In  one case, the primary variation is that of the 
seed photon temperature leading to variations of the photon density. 
This in turn will lead to a variation in the corona temperature given by
linearising Eqn. (\ref{Heat}) i.e.
\begin{align}
\frac{3}{2}kT_{eo}\Delta T_e (i\omega) &= \frac{\sigma_T c}{m_ec^2}
\left[\int 4kT_{eo}
(\Delta T_e \right. \nonumber \\ & \left. +\Delta n_{\gamma})
n_{\gamma o}\ EdE -\int E^2\Delta n_{\gamma}n_{\gamma o}dE\right]
\label{Heatlin}
\end{align}
where the heating rate of the corona has been assumed to be a constant.
 
In  the other case, the primary variation is that of  the coronal 
heating rate, i.e. $\dot{H}_{Ex} 
= \dot{H}_{Exo} (1+ \Delta \dot{H}_{Ex} \ e^{-i\omega t})$, which will lead to
the variation in the corona temperature and its amplitude is computed as: 
\begin{align}
\frac{3}{2}kT_{eo}\Delta T_e (-i\omega) &= \dot{H}_{Exo}
\Delta \dot{H}_{Ex}  -\frac{\sigma_T c}{m_ec^2}\left[\int 4kT_{eo}
(\Delta T_e \right. \nonumber \\ & \left. +\Delta n_{\gamma})
n_{\gamma o}\ EdE -\int E^2\Delta n_{\gamma}n_{\gamma o}dE\right]
\label{Heatlin2}
\end{align}
which is the linearised form of Eqn (\ref{Heat}). 
The seed photon temperature will fluctuate as response to the  fluctuation 
in $T_e$ and its amplitude $\Delta T_b$ is computed by linearising the Eqn. 
(\ref{feedback}) as:
\begin{equation}
4 \sigma (T_{bo})^4 \Delta T_b = \frac{\eta V_c}
{4 \pi a^2}
\int \frac{n_{\gamma o}}{(\tau^2+\tau)t_c}
\Delta n_{\gamma} \ E \ dE 
\label{feedbacklin}
\end{equation}
In both cases, the fluctuation of $T_b$ or $T_e$ are coupled to each 
other. In particular, their coupled amplitudes $\Delta T_e$ $\&$ $\Delta T_b$ 
are obtained
by solving  two complex equations (\ref{Komlin}), (\ref{Heatlin}) or  
(\ref{Komlin}), (\ref{feedbacklin}) simultaneously in an iterative manner.

\begin{figure*}
\includegraphics[width=0.34\textwidth]{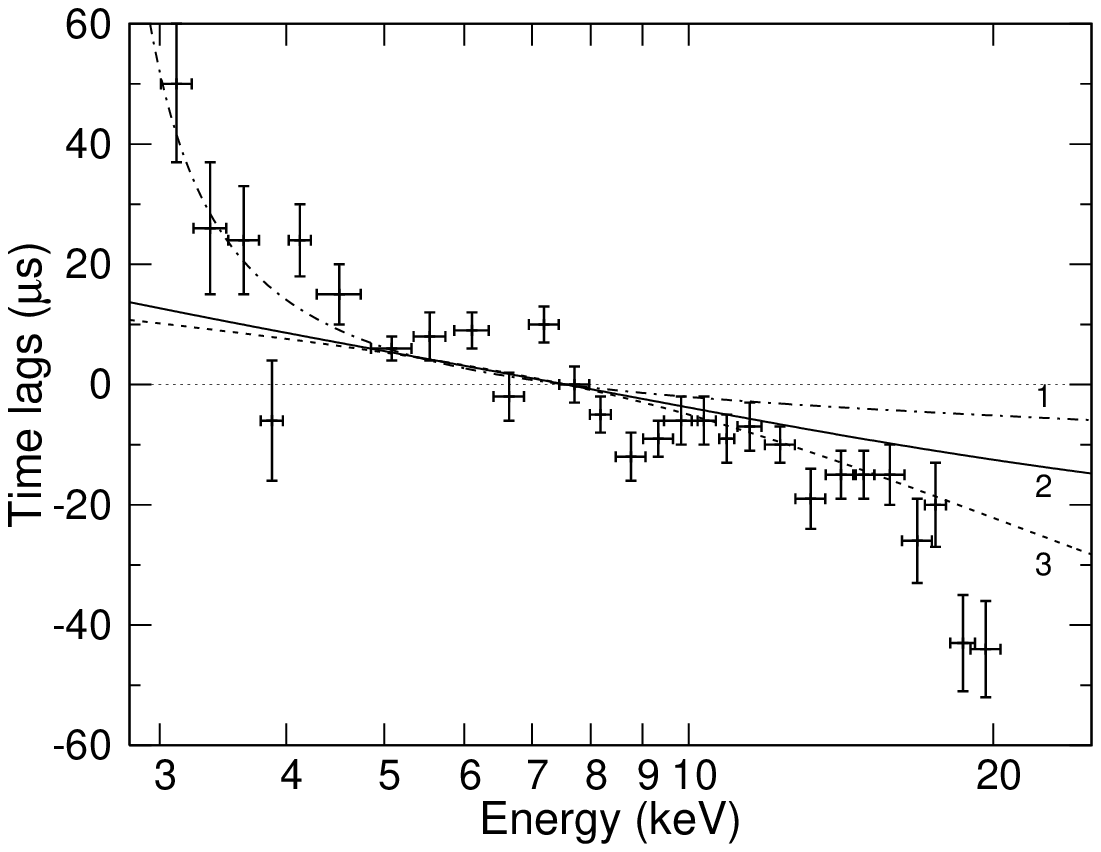}
\includegraphics[width=0.34\textwidth]{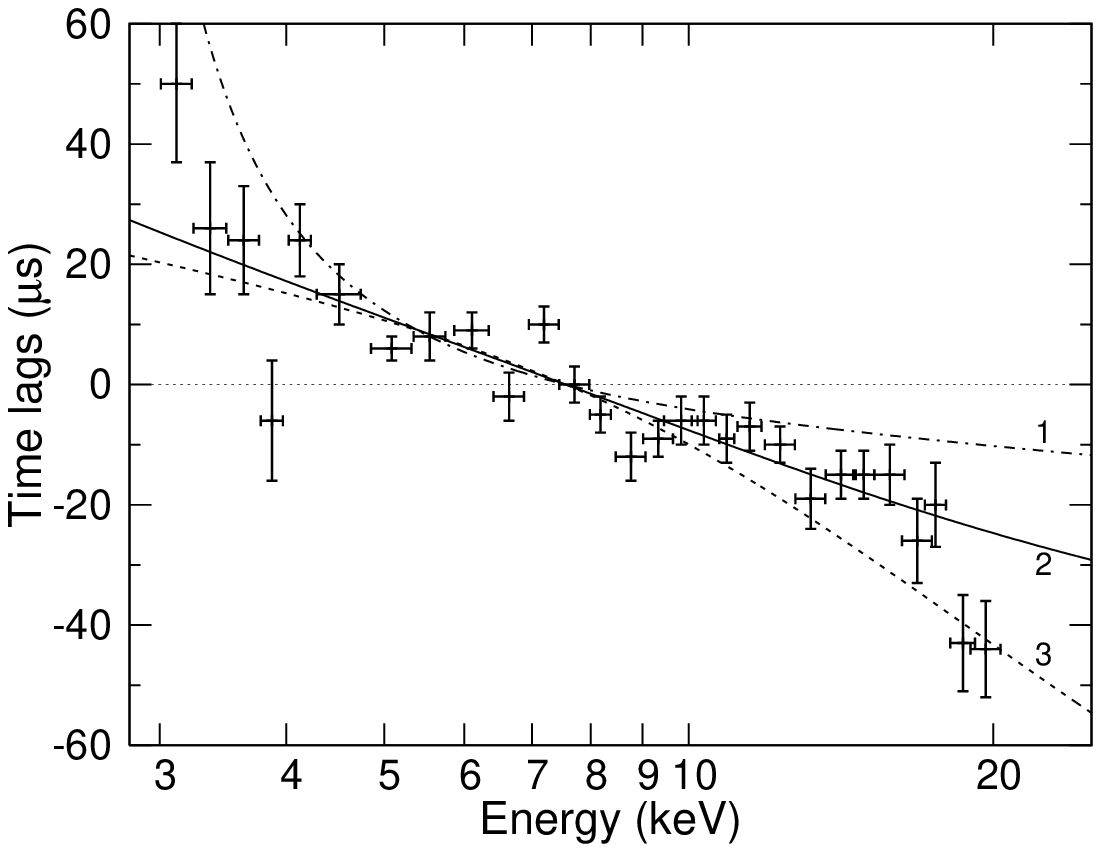}
\includegraphics[width=0.34\textwidth]{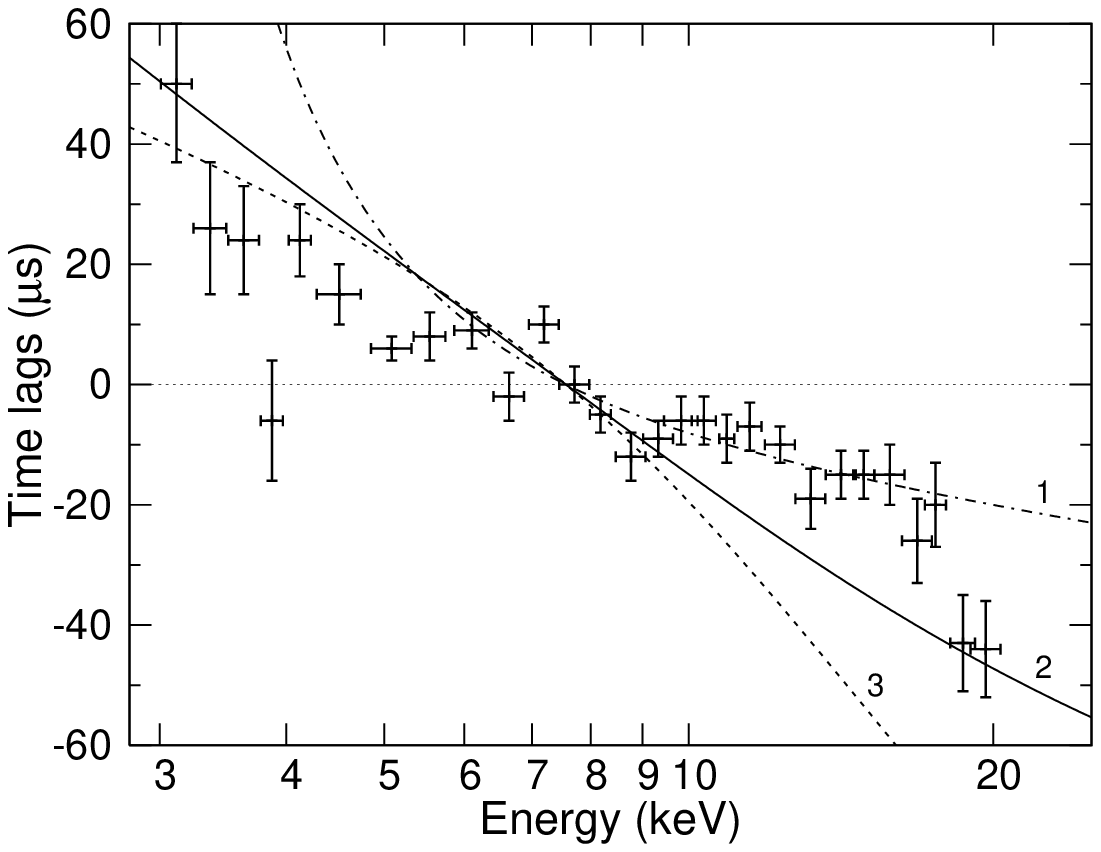}
\caption{Time lags versus energy for the 
3-March-1996 
observation of 4U 1608-52 when the time-averaged spectral model is the  
`hot' seed photon one. The 
left, middle and right panels are for L = 0.25, 0.50, and 1.0 km respectively and
the curves marked 1, 2, and 3 are for $\eta = $ 0.35, 0.55, and 0.65 respectively.
The best description of data is obtained for L = 0.5 km and $\eta$ = 0.55 (solid
line in the middle panel).
}
\label{hotlag}
\includegraphics[width=0.34\textwidth]{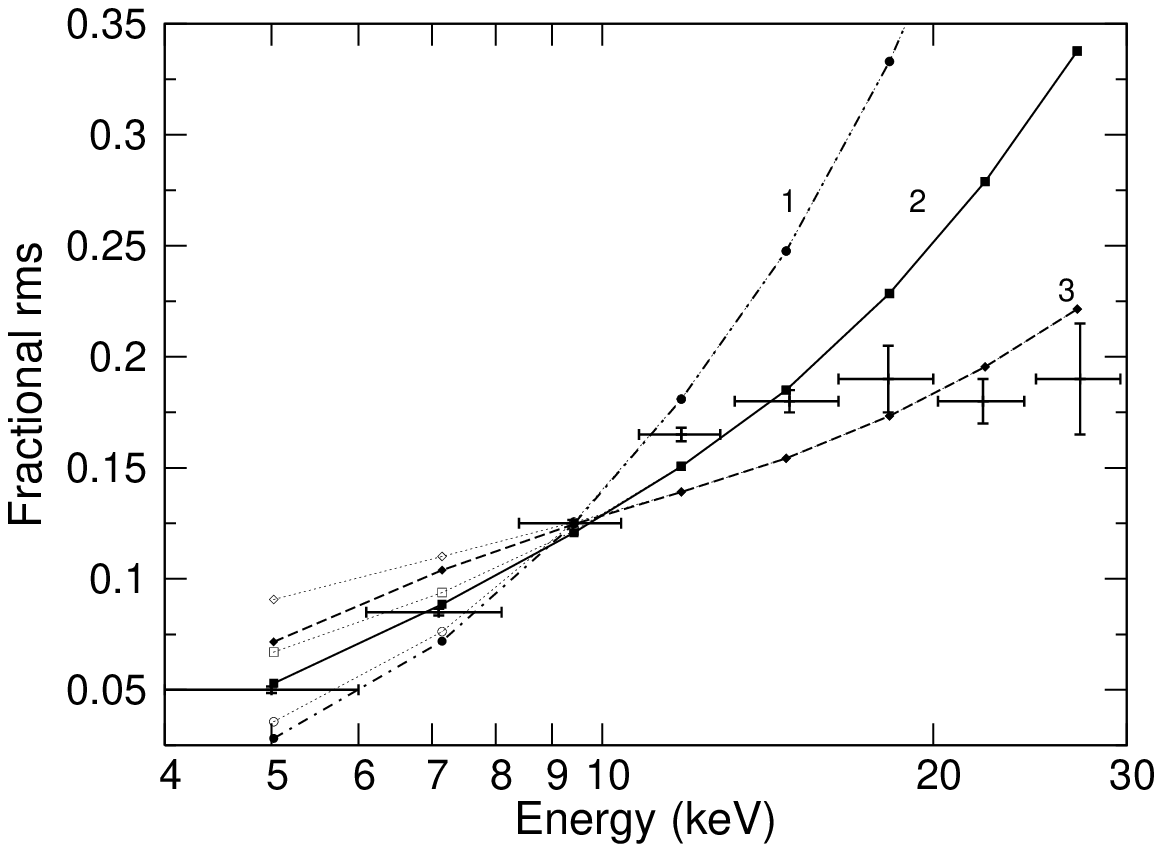}
\includegraphics[width=0.34\textwidth]{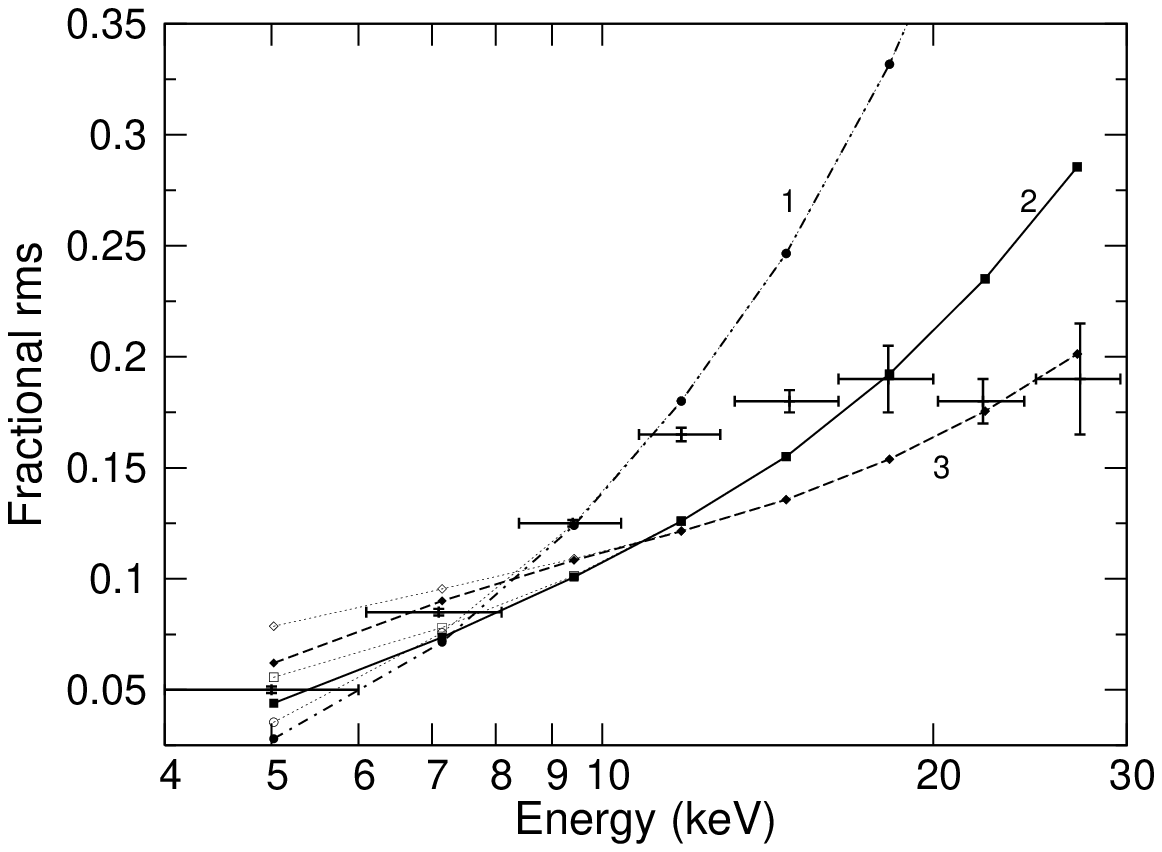}
\includegraphics[width=0.34\textwidth]{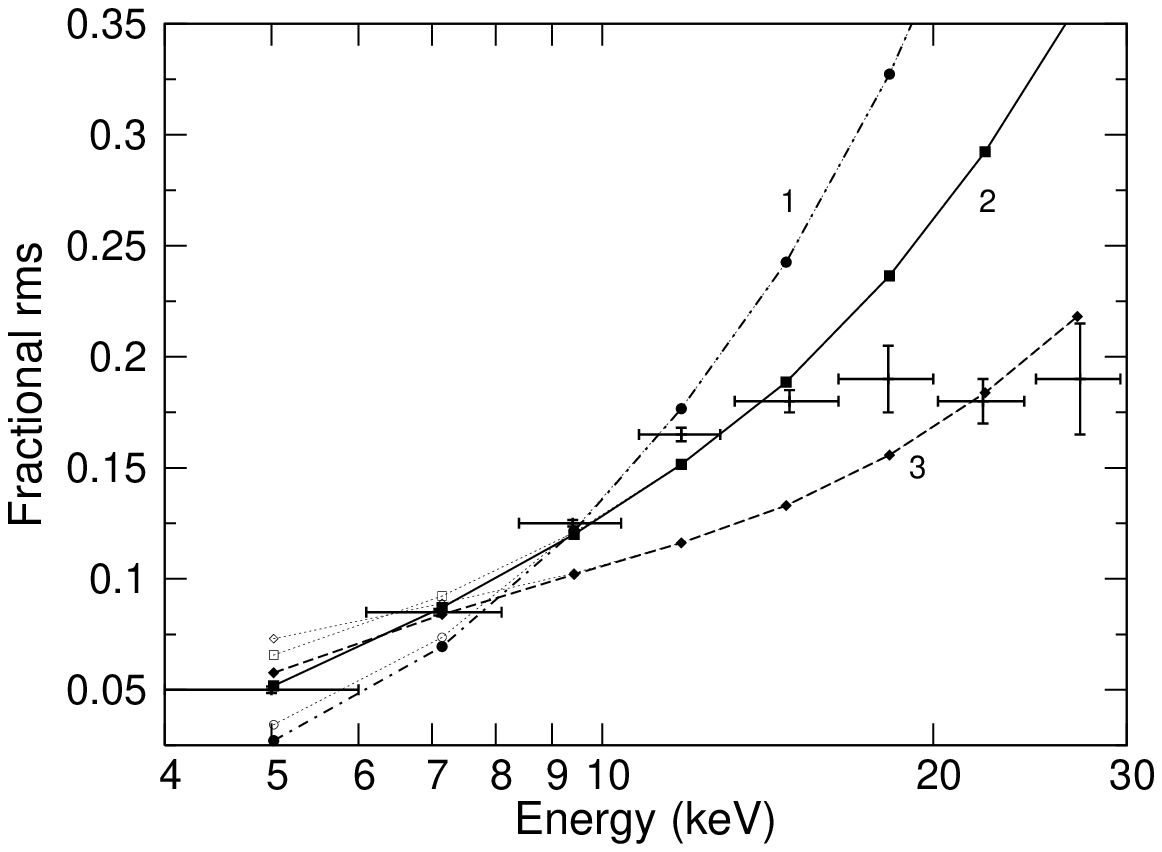}
\caption {Fractional rms versus energy for the 
3-March-1996 observation of 4U 1608-52 when the time-averaged spectral model is the   `hot' seed photon one. The predicted lines correspond to the same parameters as used for 
the lines in Figure \ref{hotlag}. The solid points refer to the r.m.s of
the main thermal Comptonization component. The open circles are the
corrected r.m.s values when the constant contributions of other components such
as the soft thermal component and the Iron line is taken into account.
}
\label{hotrms}
\end{figure*}

\begin{figure*}
\includegraphics[width=0.33\textwidth]{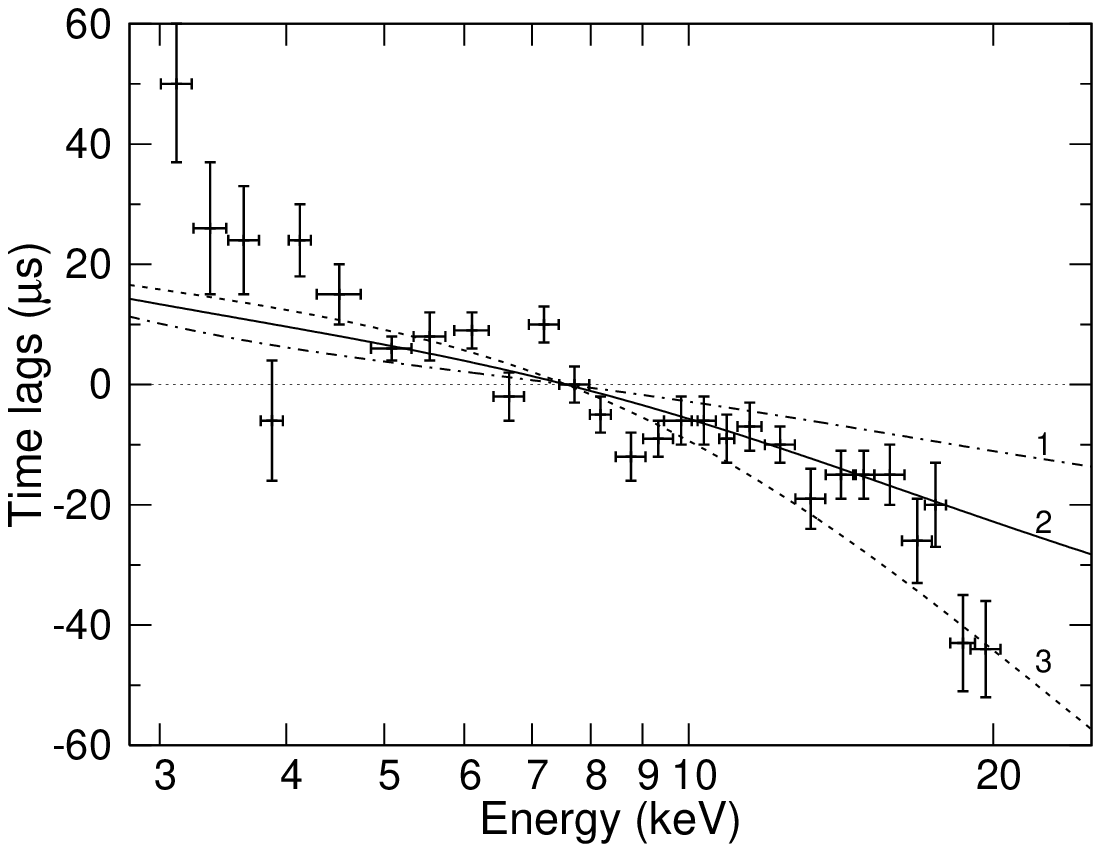}
\includegraphics[width=0.33\textwidth]{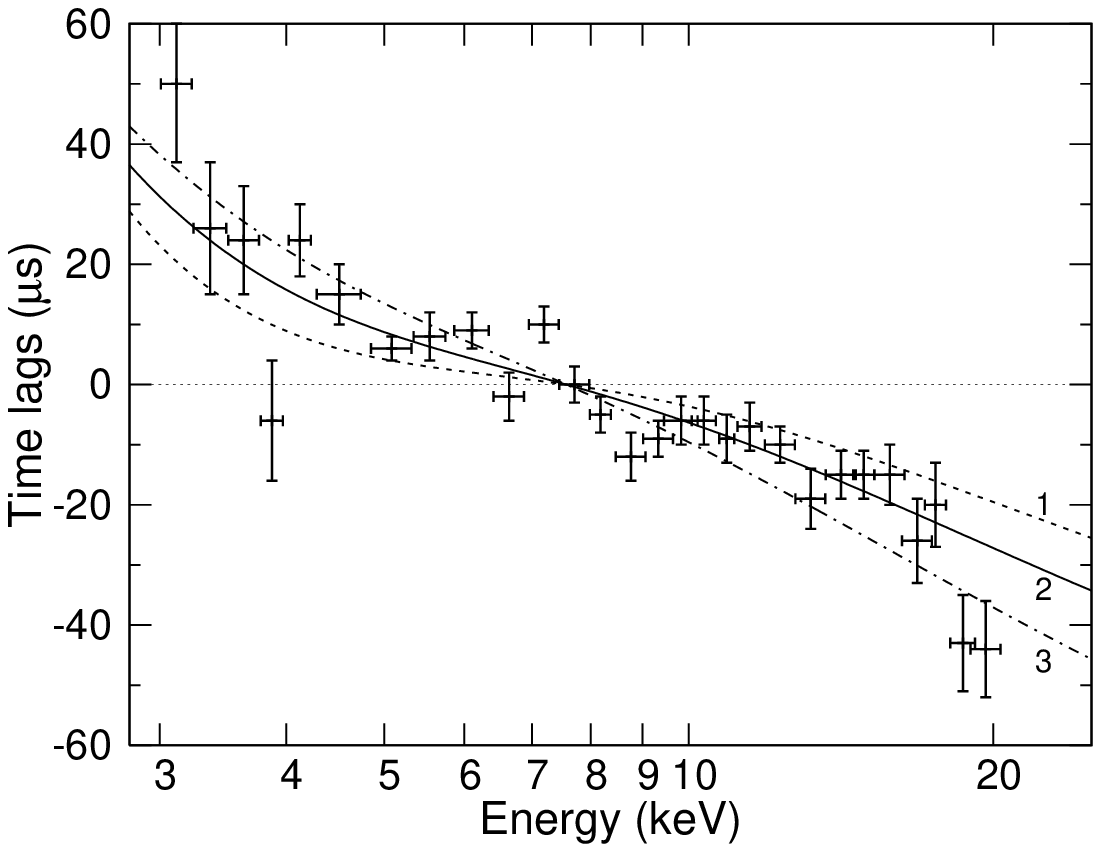}
\includegraphics[width=0.33\textwidth]{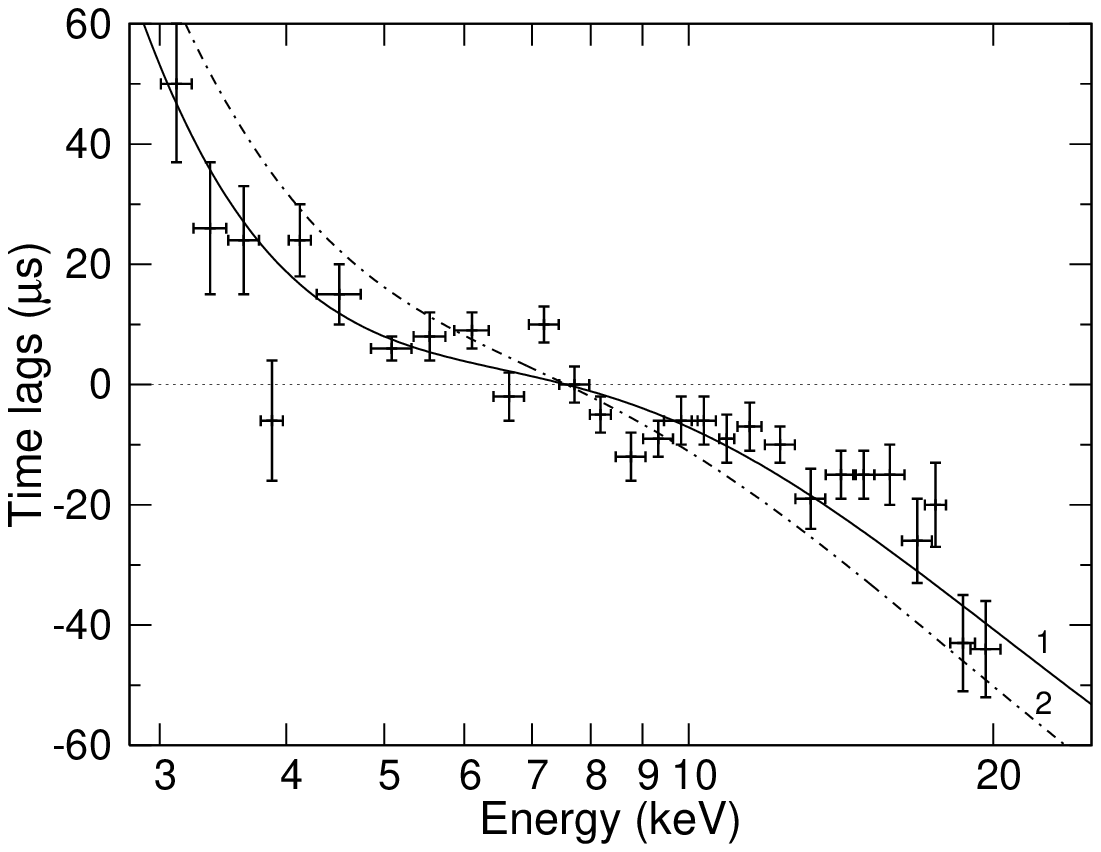}
\caption{Time lags versus energy for the 3-March-1996 
observation of 4U 1608-52 when the time-averaged spectral model is the   `cold' seed photon one. 
The left panel is for L=0.5 km and the curves marked 1, 2, and 3 
are for $\eta = $ 
0.3, 0.4, and 0.47 respectively. The middle panel is for L = 2 km and the curves 
marked
1, 2, and 3 are for $\eta = $ 0.20, 0.25, and 0.3 respectively. The right panel is for
L = 5 km and the curves marked 1 and 2 are for $\eta = $ 0.25 and 0.3 respectively.
The best description of data is obtained for L = 2 km and $\eta$ = 0.25 (solid
line in the middle panel).
}
\label{coldlag}
\includegraphics[width=0.33\textwidth]{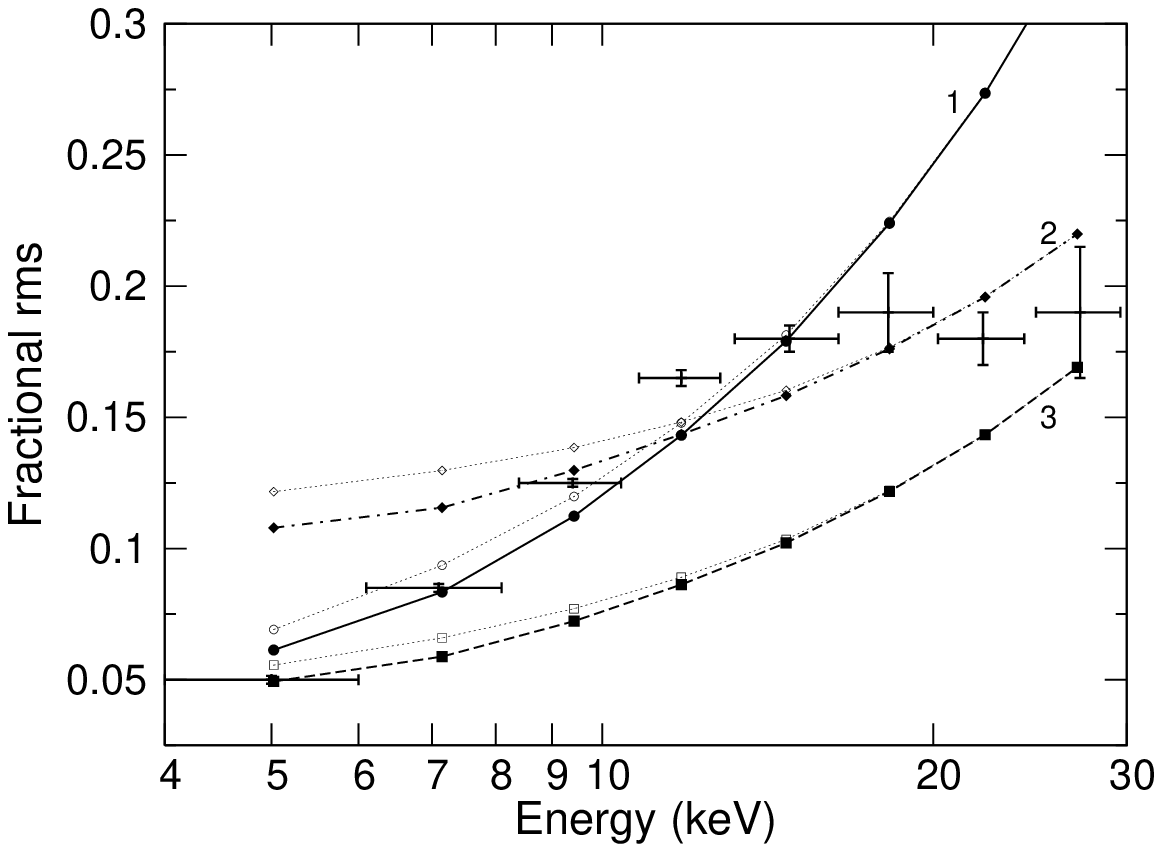}
\includegraphics[width=0.33\textwidth]{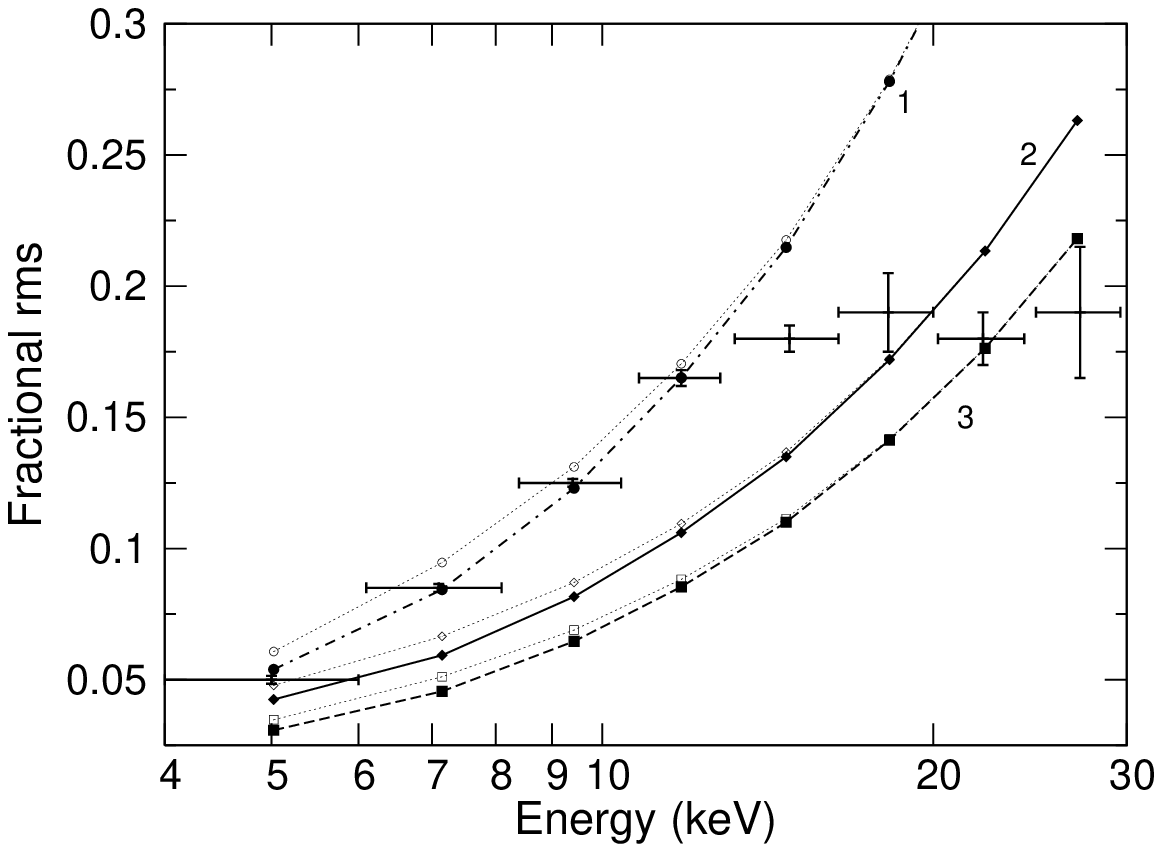}
\includegraphics[width=0.33\textwidth]{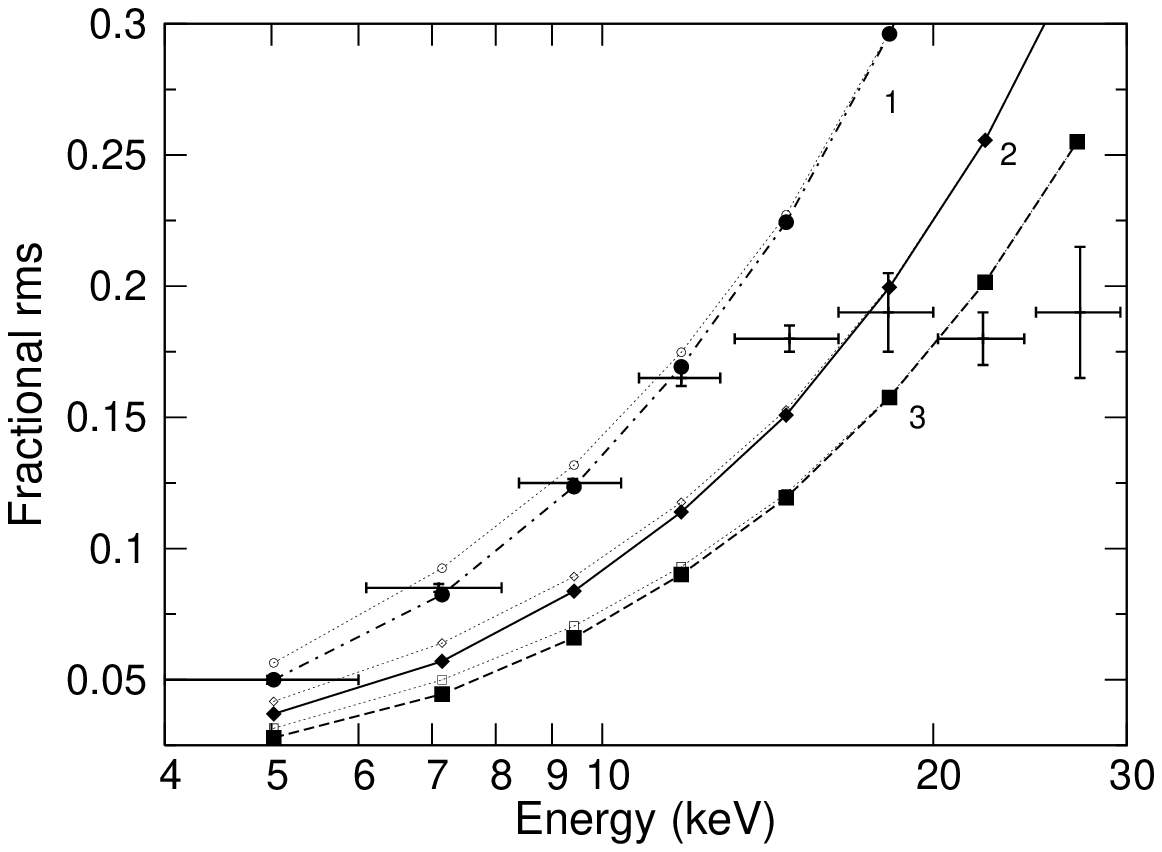}
\caption{Fractional rms versus energy for the 
3-March-1996 observation of 4U 1608-52 when the time-averaged spectral model is the   `cold' seed photon one The predicted lines correspond to the same parameters as used for 
the lines in Figure \ref{coldlag} and the points have same meaning as Figure
\ref{hotrms}. 
}
\label{coldrms}
\end{figure*}

\section{C\lowercase{omparison of \MakeUppercase X-ray spectral model with energy dependent 
temporal properties of} 4U 1608-52}

The RXTE observation of the $\sim$850 Hz QPO of 4U 1608-52 on 3rd March 1996,
remains one of the best cases for the energy dependent 
study of kHz QPOs. \citet{Berger-etal1996} studied the energy 
depended fractional r.m.s for this observation, while 
\citet{Kaaret-etal1999} and 
\citet{Vaughan-etal1997} had shown that the lower kHz QPO exhibits 
soft lag.  Recently \citet{Barret2013} have computed the
lag versus energy for a larger number of energy bins.
In Paper I, we have shown that if the oscillation  is due 
to the soft photon source, the time lags are always hard.  However,  
when the medium temperature is the driver of the oscillation, 
the time lags are  soft when a significant  fraction of the 
Comptonized photons  impinge back into the seed photon source. 
Indeed, the qualitative features of the time lag and
r.m.s versus energy were reproduced by the model and a size of the
corona was estimated for a particular set of spectral parameters.
However, 
as described in the Introduction, in general, the spectra of LMXBs can be
fitted by two spectrally degenerate models, the hot and cold seed photon
 models,  and here we study the dependence
of the time-lag and r.m.s on using these models.

\citet{Barret2013} analysed the spectrum during the
3 March 1996 observation and used a model consisting of a disk black body,
a thermal Comptonization component and an Iron emission line. The spectral parameters
obtained of the Comptonization component were  $kT_e = 2.69$ keV, optical depth 
 $\tau^2$ = 38.9 and soft photon temperature of $1.02$ keV. It was these spectral
parameters that we used in Paper I to estimate the predicted time lag and r.m.s as
a function of energy. For clarity, we reproduce here  the same results 
in Figures
\ref{hotlag} and \ref{hotrms}. The left, middle and right figures are for 
different values of
the size of the system while the curves drawn are for different values of the fraction 
impinging back to the soft source, $\eta$. The curves shown in the left and right panels of Figure
\ref{hotlag} which are for a size of 0.25 and 1 kms respectively, do not
match the observed points, while the curve in the middle panel for a size of 0.5 kms predicts 
the observed values. Hence, from these Figures one can deduce that
for this spectral model, the size of the corona should be around 0.5 kms and
$\eta \sim 0.5$. The spectral model used by \citet{Barret2013} represents
the hot-seed photon model since the seed photon temperature is high, i.e. $\sim 1$ keV.

In the literature, we did not find spectral analysis of this observation for the
cold-seed photon model. Hence,  we performed the  spectral analysis of the data
using a model consisting of a black body, a Comptonized component and an Iron line,
all modified by absorption. The fitting was done using the XSPEC package
where the model was described by wabs*(BB+CompTT+Gauss). The hydrogen column density for the  interstellar
absorption was fixed at $N_H = 1.5 \times 10^{22}
cm^{-2}$ \citep{Barret2013} and the intrinsic width of Gaussian line was
fixed to 0.1 keV \citep{Lin-Remillard-Homan2007}. 
The best fitted spectral parameter for the Comptonized component were
2.66 keV, 40.4 and 0.4 keV for electron temperature ($kT_e$), optical 
depth ($\tau^2$), and seed photons temperature ($kT_b$) respectively. 
The low value of the seed photon temperature implies that this model belongs to
the cold-seed photon family. Our motivation
here is not to describe the spectrum accurately, but rather to understand the
effect of the different spectra on the r.m.s and time lags of the kHz QPO.
Figures \ref{coldlag} and \ref{coldrms} compare the observed time-lag and r.m.s with
the predicted values for the cold-seed model. 
For both the cold and hot-seed models, there is a range of size and $\eta$ that
broadly explains the time lag and r.m.s dependence on energy and hence the 
temporal
behaviour cannot distinguish between the two. However, the range of allowed 
values, especially
for the size of the system, depends on the model used.  In the hot-seed model 
the allowed range of L size is from 0.25 to 1 km, while for the 
cold-seed model the allowed range is 0.5 to 5 km. 

There are a couple of points highlighted in Paper I, which need to be 
re-emphasised here.
The model generally  over predicts r.m.s values at high energy which is also
true when the spectral model is the cold-seed one. This suggests the 
possibility that
there may be an additional non varying high energy component. 
The other point is that
there are known additional components like the soft component (which is modelled 
as either a black body or disk black body remission) and the Iron line feature.
However, these components if assumed not to be varying at the kHz frequency, 
do not have
much effect on the energy dependence of the r.m.s. 
This is demonstrated in Figures
\ref{hotrms} and \ref{coldrms} where the filled circles represent 
when the r.m.s is
corrected for these components while the open circles are when it is not.

While we compare the predictions with the observed data, a formal fitting
of the data has not been attempted. This is because of the several assumptions
such as the geometry of the system and neglection of additional 
time delay due to reprocessing. However, we mention in passing that 
it seems the modelling prefers the hot-seed
model by providing perhaps a more realistic size for the corona and
that it seems to provide a ``better'' fit to the energy dependent time lag 
than the cold-seed photon one (compare the solid line in the middle panels
of Figures \ref{hotlag} and \ref{coldlag}).

\section{P\lowercase{ossible \MakeUppercase driver for upper k\MakeUppercase Hz \MakeUppercase{QPO}s in} 4U 1608-52}

For the same 3-March-1996 observation of 4U 1608-52, 
\citet{Mendez-etal1998a} reported the
detection of a simultaneous upper kHz QPO at $\sim$1050 Hz and have computed 
the fractional
r.m.s as a function of energy till about $\sim 20$ keV. Later by 
averaging several data sets,
\citet{Mendez-vanderKlis-Ford2001} present the fractional r.m.s for 
the upper kHz QPO and
we consider the r.m.s. for energies $> 20$ keV from this averaged analysis. 
There are no reported measure of the energy dependent time-lag for the 
upper kHz QPO
for the 3-March observations and hence we consider the time-lag computed by 
\citet{deAvellar-etal2013}
obtained by averaging several observations.

\begin{figure*}
\includegraphics[width=0.33\textwidth]{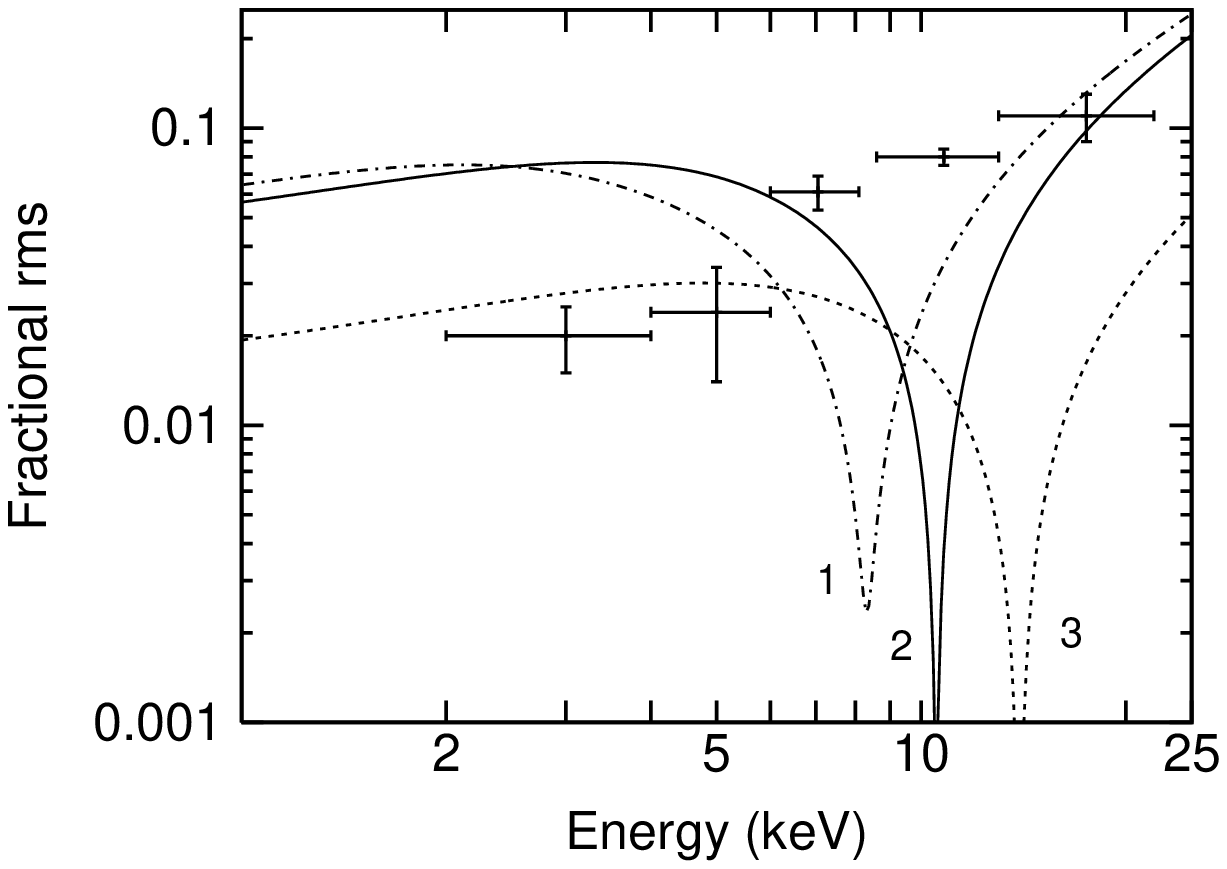}
\includegraphics[width=0.33\textwidth]{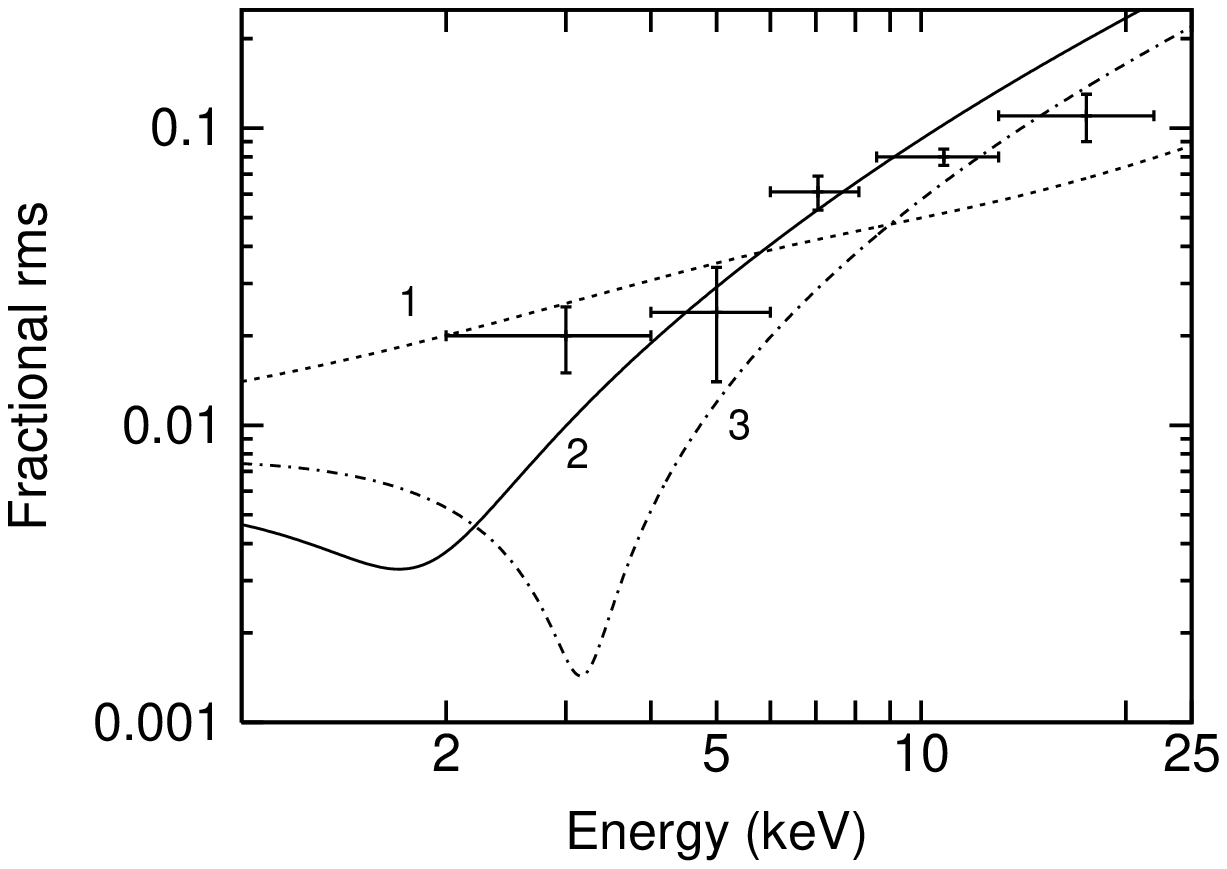}
\includegraphics[width=0.33\textwidth]{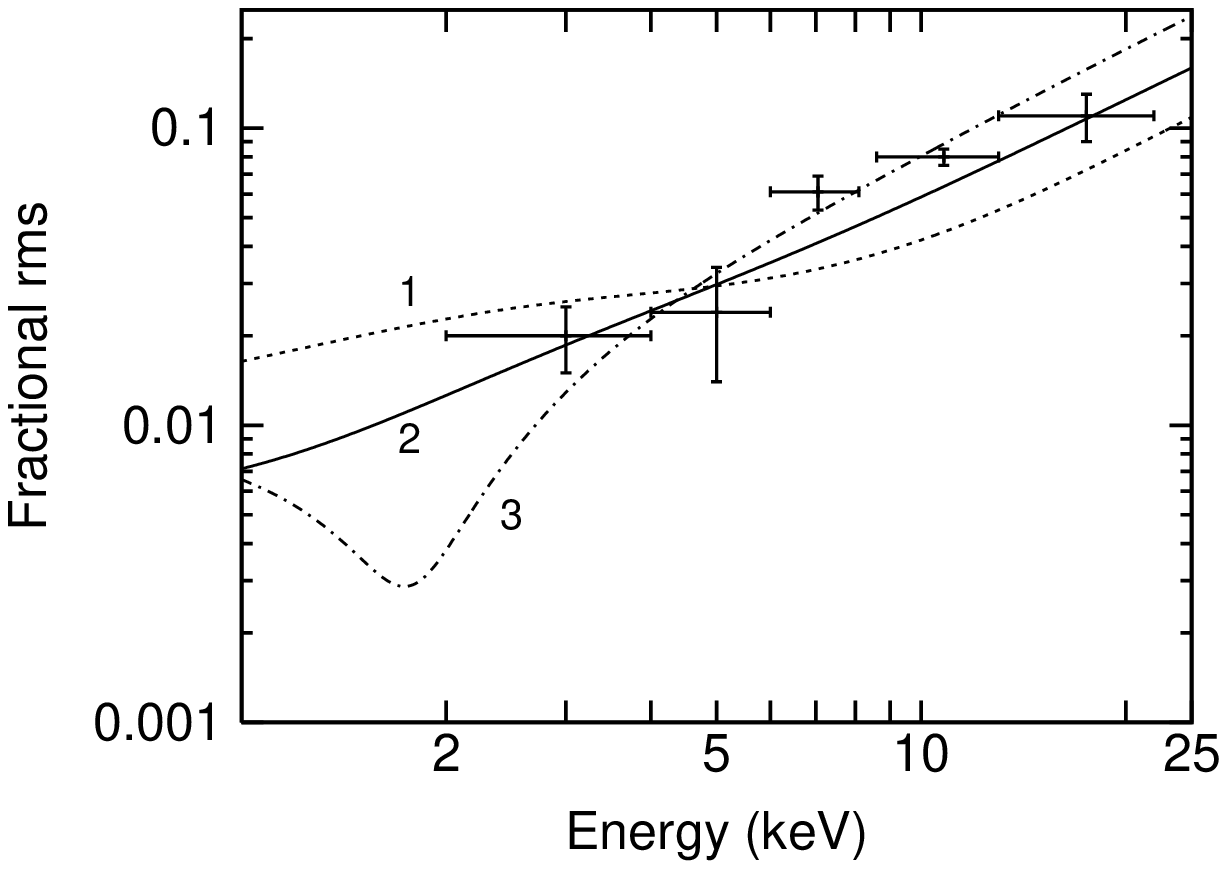}
\caption{Fractional rms versus energy for the upper
kHz QPO ($\sim 1050$ Hz) of 4U 1608-52. The first three low energy data
points are taken from the observations on 3-March-1996 and rest high energy 
data points are the values obtained by  averaging over many observations 
when the  upper KHz QPO has a frequency  $\sim 1000$ Hz. 
The left panel is for the case when the primary driver is
in the seed photon temperature.
The curves are for different time-averaged spectral models
and sizes. Curves 1 and 2  are for the cold see photon models with L = 2 
and 1 km respectively. Curve 3 is for the hot seed photon model with
L = 1 km.
The middle and right panels correspond to the case when the primary oscillation
is in the medium temperature.
The middle panel 
is for the hot-seed model with  L = 0.5 km and the curves marked 1, 2, and 3 
are for $\eta = $  0.65 (=$\eta_{max}$), 0.4, and 0.3 respectively. 
The right panel is for the cold-seed model with L = 2.0 km
and the curve marked 1, 2, and 3 are for $\eta = $ 
0.47 (=$\eta_{max}$), 0.3, and 0.15 respectively. Note that only for the
cases considered in the left panel, the associated time-lags will be hard
and for the others it would be soft  
}
\label{upperrms}
\end{figure*}

Since the lower and upper kHz QPOs are observed simultaneously, 
the geometry of the system and the time-averaged spectral parameters 
should be naturally same
for both QPOs. If like the lower kHz QPO, 
the upper one is also driven by variations
in the medium temperature, then the nature of the r.m.s and time-lag behaviours should
be similar for both of them. 
In particular the expected time-lag should be soft. However
this seems to contradict the results obtained by 
\citet{deAvellar-etal2013} where the
time-lag for the upper kHz QPO was reported to be hard or qualitatively 
different than
those of the lower kHz QPO. This suggests that the perhaps the driving 
mechanism for
the two QPOs is different. Indeed, as we had reported  
in Paper I, if the QPO is
driven by variation in the soft photons source, the time-lag 
is expected to be always
hard. However, as can be seen in Figure \ref{upperrms}, the r.m.s versus 
energy data
points are similar to the lower kHz QPO in the sense that the r.m.s increases 
with
energy. On the other hand, as shown in Paper I, if a QPO is driven by the soft
photon source the r.m.s behaviour is different and has a pivot point. 
Thus, when, we try to fit the
r.m.s of the upper QPO using such a model as shown in the left panel of
 Figure \ref{upperrms} we always find a pivot point at $\sim 10$ keV 
which is not
seen in the data.  In contrast, when we consider 
fluctuations in the corona temperature
either the hot or cold seed model (middle and right panels of Figure 
\ref{upperrms}) explains
the qualitative nature of the observed curve with perhaps the cold seed 
photon model being
more favourable.

While, the energy dependent r.m.s of the  upper kHz QPO suggests that the 
primary driver
is the coronal temperature, the time-lag suggests otherwise. 
This is a fairly serious problem
for the model, however as we discuss later in the last section, 
it will be prudent to get better and more concrete
evidence that the time-lags are indeed hard, before 
alternate ideas are explored.

\section{T\lowercase{he \MakeUppercase Comptonizing medium width $\&$ 
the k\MakeUppercase Hz
\MakeUppercase{QPO}s in 4\MakeUppercase U 1608-52}}%

Like other, NS LMXBs, 4U 1608-52  shows a  wide range of kHz QPOs frequencies 
$\sim$400 -- $\sim$1200Hz. Unfortunately, 
unlike the March 3rd observation described in the 
previous section, there are no other reported observations with similar 
high quality. Using the {\it RXTE} observations of the
1998 outburst of 4U 1608-52,
\cite{Barret2013} found several observations which have kHz QPOs 
between $\sim$560 and $\sim$810 Hz. For some of these they  measured 
the time-lag between two broad energy bands  (e.g. 3-8 and 8-20 keV)
In this work, we refer to these values as mean lags.
We have chosen 9 observations from their data for which the observed 
average kHz QPO frequencies ranged from 580 to 850 Hz.   
For each observation, we fit  the spectrum, 
as described in \S 3, using both the hot-seed and cold-seed  spectral models.
The observations IDs and the corresponding kHz QPO frequency along with 
thermal Comptonization spectral parameters are listed in Table \ref{qpoID}.  
The total r.m.s. in the energy band 2-60 keV for these lower kHz QPOs 
have been measured
by \cite{Mendez-vanderKlis-Ford2001}. The mean soft lag, total r.m.s. and
$\eta_{max}$ for each
kHz QPO  are listed in Table \ref{qpolag}. 
Here,  $\eta_{max}$, for both hot $\&$ cold-seed spectral models, is computed
using Eqn [\ref{feedback}].

\begin{figure}
\vspace{-2.5cm}\hspace{-1.15cm}
\includegraphics[width=0.6\textwidth]{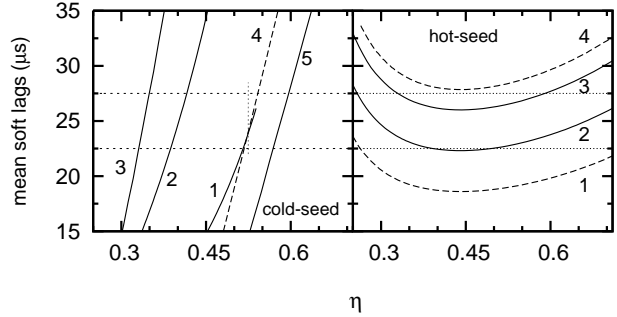}
\vspace{-1.cm}
\caption{The computed mean soft lags as a function of $\eta$ at a given size 
L. The left panel is
for the cold-seed photon model and the curves 1, 2, 3, 4, and 5 are for L = 0.6, 1.5, 5, 11, and 12 km 
respectively. The right panel is for the hot-seed photon models and the 
curves 1, 2, 3, and 4 are
for L = 0.5, 0.6, 0.7, and 0.75 km respectively. The dashed horizontal lines
mark the observed range of mean soft lag.
}
\label{hc_slt}
\end{figure}

\begin{figure}
\vspace{-2.5cm}
\includegraphics[width=0.55\textwidth]{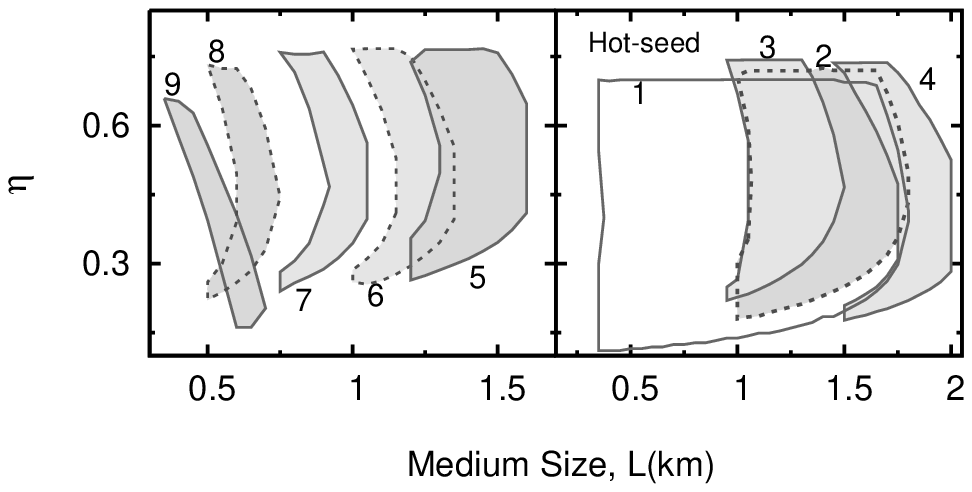}\vspace{-2.85cm}
\includegraphics[width=0.55\textwidth]{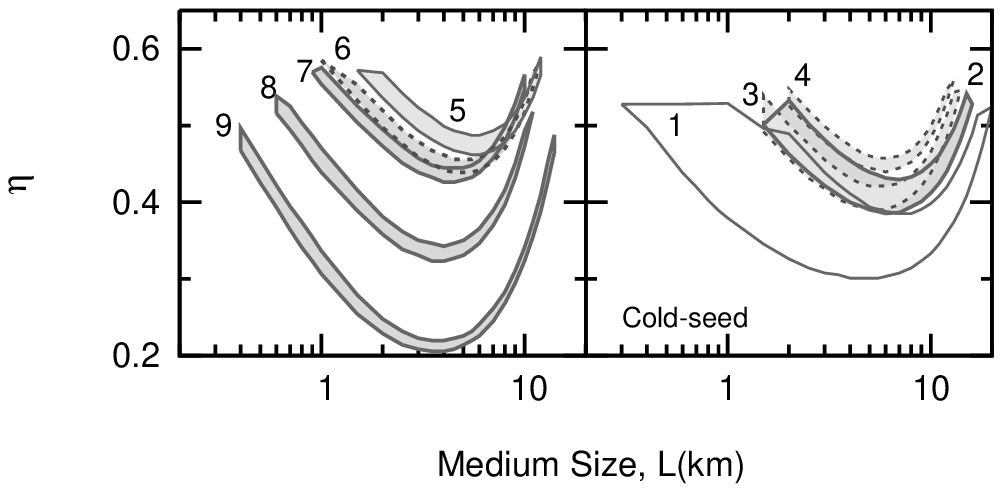}
\vspace{-1.cm}
\caption{The computed  allowed ranges of $\eta$ and medium size, L. The upper
panels correspond to the hot seed photon model while the lower ones
correspond to the cold seed photon one.
The closed curves 9, 8, 7, 6, 5, 4, 3, 2, and 1 are
for lower kHz QPO frequencies of 850, 810, 770, 740, 710, 680, 650, 615, and 580 Hz respectively.
}
\label{eta_l}
\end{figure}

\begin{figure*}\hspace{-0.6cm}
\includegraphics[width=0.34\textwidth]{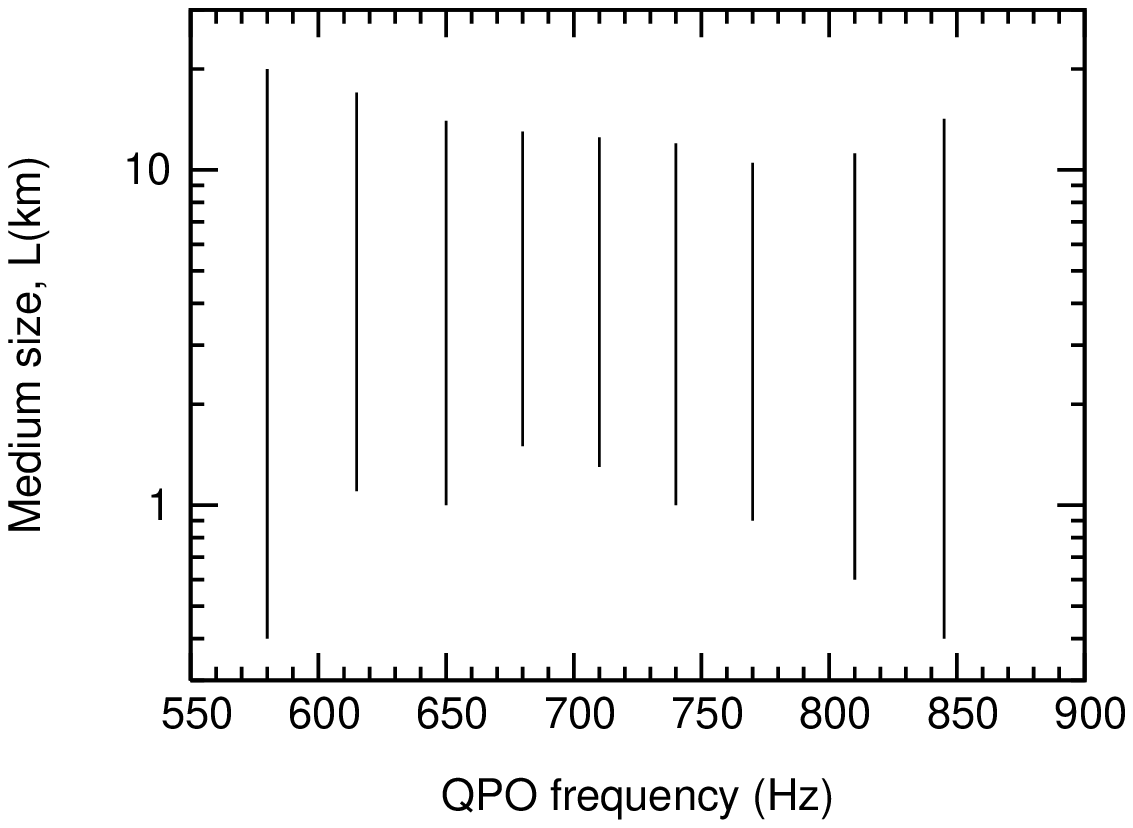}\hspace{-0.2cm}
\includegraphics[width=0.34\textwidth]{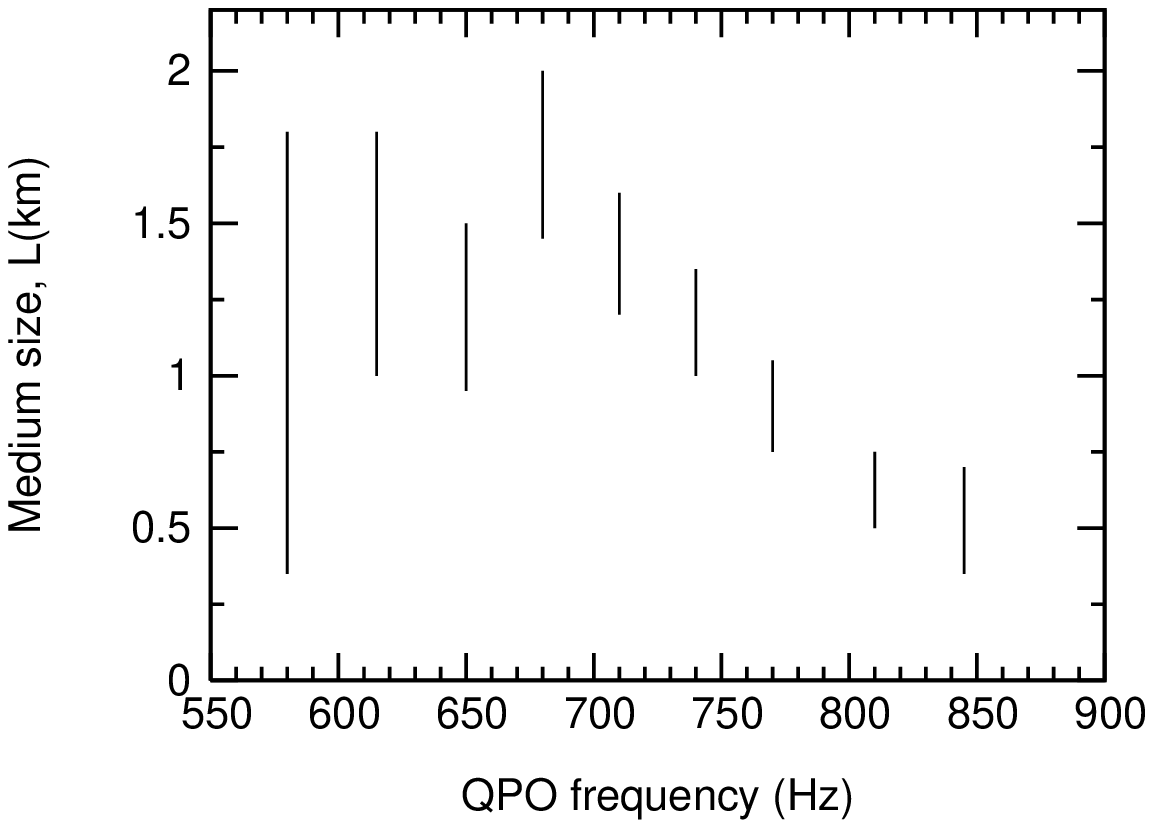}\hspace{-0.6cm}
\includegraphics[width=0.34\textwidth]{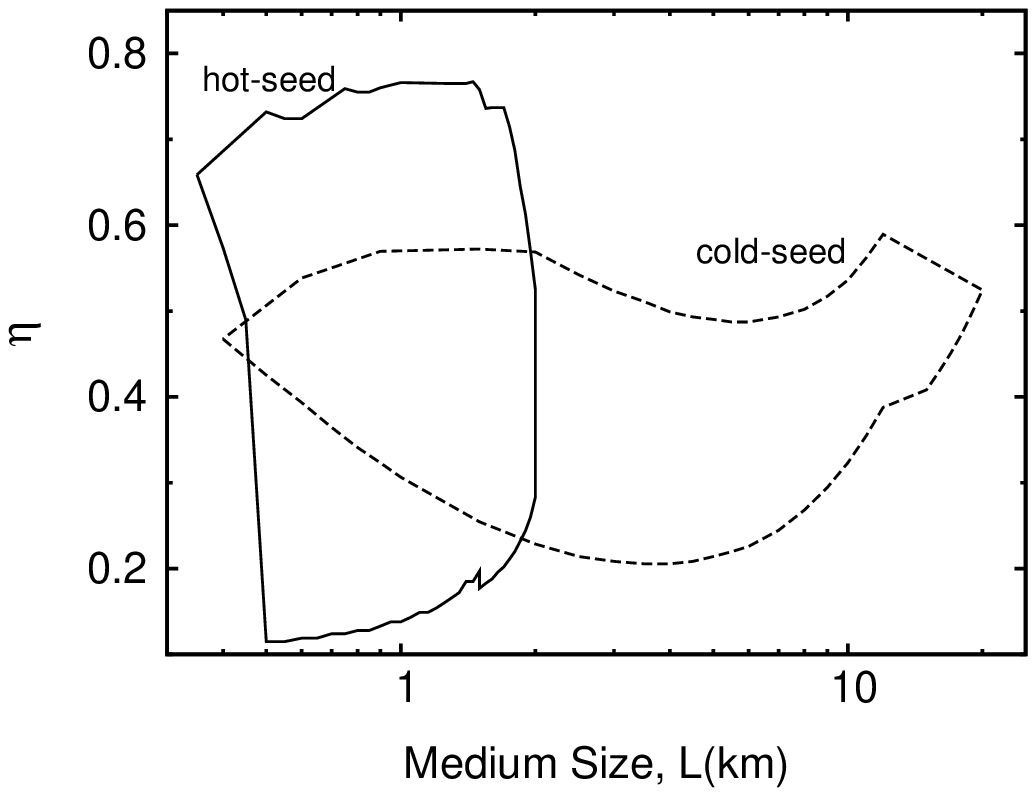}
\caption{The estimated range of size, L for the nine lower kHz QPO frequency. 
The left panel is for the cold-seed photon model while the middle is for the 
hot-seed photon one.  The right panel represents the  allowed 
 region of $\eta$ and L by overlapping the regions for each individual
frequencies shown in Figure \ref{eta_l}.
}
\label{hot_col_lrange}
\end{figure*}

We now attempt to constrain the medium size and $\eta$ for each of these
observations. Since only the mean time-lag is available, we use the following
procedure. We compute the expected mean time lag as function of $\eta$
for different size or widths of the medium. As an example, the right panel of  
Figure \ref{hc_slt}, shows the result of these computation for the
 observation corresponding to a kHz frequency of 810 Hz using the hot seed
photon spectral parameters. The mean time
lag as a function of $\eta$ is a non-monotonic curve with a minimum and the
lags increase with increasing size. The two dashed horizontal lines represent
the range of the observed mean time lag and from the Figure, we can 
estimate the range of $\eta$ and width for which the curves lies within this
range. For the cold seed photon model parameters, the behaviour is more complex.
The computed mean time lag increases with $\eta$ monotonically as shown in
the left panel of Figure \ref{hc_slt}. However, the variation with width
is non monotonous with the mean time lag increasing and then decreasing
as is evident from the curves marked 1 to 5. In this case, as well we
can obtain a range of width and $\eta$ by considering only values of
$\eta$ smaller than the maximum allowed. We follow the same procedure for
the other eight observations and obtain ranges of values for  $\eta$ and
width which are plotted in Figure \ref{eta_l}. The consolidated results
of this analysis are shown in Figure \ref{hot_col_lrange} where in the
left and middle panels, the estimated
range of sizes are plotted versus the QPO frequency and the right panel
shows the overlapped allowed ranges for $\eta$ and size for 
all nine observations.
The primary results of this analysis is that for the hot seed photon model,
the size of the Comptonizing medium seems to decrease with increasing 
QPO frequency while for the cold seed photon model, no such trend is seen,
probably because the inferred size ranges are large.

\begin{table*}
\caption{
The observation IDs, the date, the frequency range and the spectral parameters 
for nine observations of 4U 1608-52.
}
\label{data_spec}
\begin{tabular}{llllllllll}
\hline \hline
ObsID & Start date/time & Freq* (Hz) & Freq (Hz) & \multicolumn{3}{l}{Hot-seed-photon model} & \multicolumn{3}{l}{Cold-seed-photon model}\\
\cline{5-10}
& & & &  \multicolumn{1}{l}{$kT_b$} & $kT_e$ & $\tau^2$ & \multicolumn{1}{l}{$kT_b$} & $kT_e$ & $\tau^2$\\
\hline
10072-05-01-00 & 1996/03/03 & 840--890 & 850 & 1.02$^{+0.10}_{-0.10}$ & 2.69$^{+0.04}_{-0.03}$ & 38.81$^{+2.08}_{-2.02}$ & $0.40^{+0.10}_{-0.40}$ & 2.6$^{+0.09}_{-0.06}$ & 40.47$^{+3.72}_{-4.48}$ \\
30062-02-01-000 & 1998/03/24-17:06:23 & 796-819 & 810 & 1.23$^{+0.06}_{-0.06}$ & 2.92$^{+0.11}_{-0.08}$ & 25.87$^{+2.75}_{-3.03}$ & 0.36$^{+0.10}_{-0.31}$ & 3.13$^{+0.36}_{-0.24}$ & 24.33$^{+5.72}_{-5.68}$ \\
30062-01-01-00  & 1998/03/27-15:29:21 & 752-795 & 770 & 1.28$^{+0.04}_{-0.04}$ & 3.37$^{+0.18}_{-0.14}$ & 15.66$^{+2.15}_{-2.17}$ & 0.34$^{+0.09}_{-0.23}$ & 4.10$^{+0.82}_{-0.61}$ & 13.72$^{+5.30}_{-5.14}$\\
30062-01-01-00  & 1998/03/27-13:53:23 & 723-750 & 740 & 1.28$^{+0.04}_{-0.04}$ & 4.28$^{+0.59}_{-0.39}$ & 9.03$^{+2.44}_{-2.37}$ & 0.35$^{+0.07}_{-0.28}$ & 4.87$^{+1.11}_{-0.64}$ & 10.46$^{+3.18}_{-3.17}$\\
30062-01-01-00  & 1998/03/27-12:27:36 & 708-743 & 710 & 1.31$^{+0.03}_{-0.03}$ & 4.65$^{+0.92}_{-0.50}$ & 7.89$^{+2.43}_{-2.66}$ & 0.34$^{+0.07}_{-0.29}$ & 5.98$^{+3.47}_{-1.15}$ & 7.70$^{+3.54}_{-4.22}$\\
30062-01-01-02  & 1998/03/29-10:41:15 & 661-697 & 680 & 1.20$^{+0.02}_{-0.02}$ & 5.00$^{f}$           & 7.83$^{+0.12}_{-0.12}$ & 0.34$^{+0.07}_{-0.33}$ &5.54$^{+2.86}_{-1.05}$ & 9.56$^{+4.24}_{-4.86}$\\
30062-02-01-00 & 1998/03/24-22:01:31 & 643-667 & 650  & 1.26$^{+0.03}_{-0.03}$ & 4.41$^{+0.29}_{-0.23}$ & 10.04$^{+1.35}_{-1.40}$ & 0.35$^{+0.08}_{-0.34}$ &4.68$^{+0.77}_{-0.49}$ &  12.23$^{+2.91}_{-2.99}$\\
30062-02-01-01 & 1998/03/25-18:42:22 & 595-631 & 615  & 1.23$^{+0.03}_{-0.03}$ & 5.96$^{+0.96}_{-0.69}$ & 6.32$^{+1.87}_{-1.67}$ & 0.33$^{+0.07}_{-0.26}$ & 6.49$^{+5.56}_{-1.59}$ & 7.8$^{+4.72}_{-6.34}$\\
30062-02-01-01 & 1998/03/25-17:46:60 & 567-576 & 580 & 1.17$^{+0.04}_{-0.03}$ & 7.25$^{+3.98}_{-1.05}$ & 5.08$^{+1.80}_{-2.95}$ & 0.34$^{+0.06}_{-0.33}$ & 6.40$^{+1.28}_{-0.75}$ & 8.6$^{+2.03}_{-2.26}$\\
\hline
\end{tabular}\\
\label{qpoID}
Note: Freq* is the observed frequency range  for the observation 
while Freq  is the frequency used to compute time-lags. $kT_b$ is the 
seed photon source temperature, $kT_e$ the   
 temperature of the Comptonizing medium, and $\tau$ is its optical depth.
 Here $\tau$ is different from $\tau_o$ of COMPTT and it
is calculated by the relation $\tau^2+\tau$= $\frac{12}{\pi^2}$ $(\tau_o+2/3)^2$. The observed QPO frequency range  and the
spectral parameters for the hot-seed-photon model are taken from \citet{Barret2013}
(and reference within). 

\end{table*}

\begin{table}\vspace{-0.0cm} 
\caption{The measured mean soft lag (soft lags between (3--8) and (8--30) keV
photons), the total r.m.s. (in 2--60 keV) and the calculated $\eta_{max}$ for
the observations listed in Table \ref{qpoID}}
\label{tlag-lmax}
\begin{tabular}{lllll}
\hline \hline
Freq (Hz) & Mean soft lag ($\mu s$) & Total rms ($\%$) & $\eta_{max}$ $^1$ & 
$\eta_{max}$ $^2$\\
\hline
850 & 20.5 $\pm$2.0 & 7.5 $\pm$0.5 & 0.655 & 0.476\\
810 & 25.0 $\pm$2.5 & 8.1 $\pm$0.5 & 0.728 & 0.525\\
770 & 30.0 $\pm$2.5 & 8.3 $\pm$0.7 &0.759 & 0.569\\
740 & 30.0 $\pm$2.5 & 9.15$\pm$0.65 & 0.767 & 0.581\\
710 & 34.0 $\pm$4.0 & 9.2 $\pm$0.8 & 0.766 & 0.587\\
680 & 39.0 $\pm$3.0 & 9.5 $\pm$0.7 & 0.737 & 0.555\\
650 & 31.0 $\pm$5.0 & 8.8 $\pm$1.0 & 0.740 & 0.553\\
615 & 28.0 $\pm$7.0 & 7.6 $\pm$0.6 & 0.724 & 0.548\\
580 & 18.0 $\pm$12.0 & 6.3 $\pm$0.9 & 0.700 & 0.528\\
\hline
\end{tabular}\\
Note: The data points for the mean soft lags are taken from 
\citet{Barret2013} and for the total r.m.s from \citet{Mendez-vanderKlis-Ford2001}.
The superscript 1 and 2 are for when  $\eta_{max}$ is computed 
for the hot-seed and cold seed models.
\label{qpolag}
\end{table}

\section{Summary and Discussion}

In Paper I, we had shown that the observed soft lags of the lower 
 kHz QPO of neutron star systems, can be explained within the framework
of a thermal Comptonization model. Soft lags are expected if the
driving oscillation causes variation in the temperature of the medium and
if a certain fraction of the photons impinge back to the soft photon source.
Apart from identifying the radiative process for the oscillation, the model
also holds the promise to give an estimate for the size and geometry of
the Comptonizing medium.

In this work, we study the dependence of the results on the  model used 
to represent the time averaged spectrum during the oscillation. In the
3-20 keV spectral band of the RXTE's PCA, low mass X-ray binaries usually allow
for two spectrally degenerate models which are termed as the ``hot'' and ``cold''
seed photon model. We also study the possible variation of the size of the
medium as a function of the QPO frequency and attempt to extend the model
to data from the upper kHz QPO.

The 3rd March 1996 observation of 4U1608-52 provides one of the best
quality time-lag versus energy data for the lower kHz QPO. We compare
the  model predictions for both the hot and cold seed photon spectral models. 
We find that qualitatively, for both these spectral shapes, the Comptonization model can explain the r.m.s and time lag as a function of energy. However,
the range of the medium size inferred from the hot seed photon model, 
$0.25$-$1.0$ kms is different from that of the cold seed photon model,
 $0.5$-$10.0$ kms. Both interpretations require a significant fraction of
the photons to impinge back into the soft source with $\eta > 0.2$. 
While other observations of 4U1608-52 are not of the same high quality as
the 3rd March 1996 observations, \cite{Barret2013} have measured for
some of them lag between two broad energy bands (e.g. 3-8 and 8-20 keV).
For nine such observations corresponding to different QPO frequencies 
we have attempted to estimate the medium size using both the hot and
cold seed photon time-averaged spectral models. For the hot seed photon
model, the size of the medium is found to decrease with increasing QPO frequency, while no such trend can be inferred for the cold seed photon model case, 
probably because here the allowed range for the sizes are larger.

These results show that while it is promising to obtain the size and
geometry of the Comptonizing corona by modelling the energy dependent time-lags,
an accurate non-degenerate spectral model for the time-averaged spectrum is
critical for such an analysis. Unfortunately, the 3-20 keV energy band
of the RXTE PCA does not seem to be sufficient to do this. What is required
is a wider band spectral coverage combined with high frequency timing
capabilities. The recently launched satellite {\it ASTROSAT}\footnote{http://astrosat.iucaa.in} \citep{Agrawal2006, Singh-etal2014} will provide such a 
unique opportunity, since while the {\it Large Area X-ray Proportional counters (LAXPC)} will provide high frequency timing capability and spectra in the 3-80 keV band, the {\it soft X-ray telescope (SXT)} will provide the much needed simultaneous low energy spectra in the 0.3-8 keV band.   

During the 3rd March observation of 4U1608-52, an upper kHz QPO ($\sim 1050$ Hz)
was also observed and its fractional r.m.s versus energy is similar to the
lower kHz one. Indeed, modelling the r.m.s variation requires that the
primary driver is the medium temperature variation similar to the lower kHz QPO. However, this implies that for this QPO as well the time-lag should be soft, which seems to be in contradiction with the results obtained by 
\citet{deAvellar-etal2013} where they measure the time lag for the
 upper kHz QPO to be hard. While the significance of the lag detection is
less than that of the lower kHz QPO, this is a challenge to the simple
Comptonization model presented here. This may indicate that reverberation lags
due to time delays between a primary and its reflected spectrum maybe dominant.
While it seems unlikely that these reverberation lags can explain the
entire energy behaviour of the lower kHz QPO \citep{Cac16}, a combination
of Comptonization and reverberation lags maybe possible.
It could also be that the time averaged spectra of these sources are more complex
than assumed here and perhaps the dominant Comptonization component is not
a single temperature one but instead forms from multiple zones. Such 
complexities in the spectra including the presence of 
dominant reflection components, would require broad band sensitive instruments.

Soft lags within the thermal Comptonization framework require that a
significant fraction of the photons impinge back to the soft photon
source which  as demonstrated in this work, can be broadly constrained. 
This naturally puts restrictions on the geometry of the system.
For example, if the Comptonizing medium is a thin shell around the soft
photon generating neutron star surface, the fraction is expected to be large
compared to a geometry where the medium is hot torus like disk surrounded
by a soft photon generating accretion disk. The fraction of the
photons impinging back into the soft photon source can be estimated
for a particular geometry using a Monte Carlo code to simulate the
Comptonization process. Thus, in a near future work, 
we plan to develop and use such a Monte Carlo code to understand what
kind of geometry is allowed.

\section*{Acknowledgements}
NK thanks CSIR/UGC for providing support for this work.

\label{lastpage}

\end{document}